\documentclass[11pt]{article}

\pdfoutput=1

\usepackage{amsmath,amssymb}

\usepackage{epsfig}
\usepackage{graphicx}               % Standard graphics package
\usepackage{url}
\usepackage{hyperref}
\usepackage{cite}

\newcommand{\nc}{\newcommand}

\nc{\beq}{\begin{equation}}
\nc{\eeq}{\end{equation}}
\nc{\beqa}{\begin{eqnarray}}
\nc{\eeqa}{\end{eqnarray}}
\nc{\bea}{\begin{eqnarray}}
\nc{\eea}{\end{eqnarray}}
\nc{\ra}{\rightarrow}
\nc{\lsim}{\begin{array}{c}\,\sim\vspace{-21pt}\\< \end{array}}
\nc{\gsim}{\begin{array}{c}\sim\vspace{-21pt}\\> \end{array}}
\nc{\Tr}{{\rm Tr}}
\nc{\slsh}{\slash\hspace*{-0.22cm}}

\def\be{\begin{equation}}
\def\ee{\end{equation}}
\def\bea{\begin{eqnarray}}
\def\eea{\end{eqnarray}}
\def\bit{\begin{itemize}}
\def\eit{\end{itemize}}

\newcommand{\met}{E^{\rm miss}_T}

\def\to{\rightarrow}

\title{
\Large
\textbf{
A Toolkit of the Stop Search via the Chargino Decay
 \\
}\vspace*{1.0cm}
}

\author{Yang Bai$^{a}$, Hsin-Chia Cheng$^{b}$, Jason Gallicchio$^{b,c}$, and Jiayin Gu$^{b}$
\vspace{5mm}
\\
$^{a}$ \normalsize\emph{Department of Physics, University of Wisconsin, Madison, WI 53706, USA} \\
$^{b}$ \normalsize\emph{Department of Physics, University of California, Davis, CA 95616, USA} \\
$^{c}$ \normalsize\emph{South Pole Station, PSC 468 Box 400, APO AP 96598, Antarctica}
}

\date{}

\begin{document}
\setcounter{page}{0}
\maketitle

\vspace*{1cm}
\begin{abstract}
The top squark (stop) may dominantly decay to a bottom quark and a chargino if the mass difference between the stop and the lightest neutralino is comparable or less than the top quark mass. Such a moderately compressed spectrum is a challenging scenario for the stop search at the Large Hadron Collider, because it is difficult to separate the signals from the top and anti-top background. In this paper we focus on the di-leptonic decay channel, and consider many kinematic variables as possible discriminators. These include several $M_{T2}$ variables and new ``compatible-masses" variables which fully utilize all kinematic information of the background. We use several sample spectra with different characteristics to study the efficiencies of these variables in distinguishing the signal from the background. The finding is that different combinations of variables or strategies should be used for different spectra to maximally enhance the signal significance and expand the reach of the stop search in this scenario. The new variables that we proposed in this paper are also useful for other new physics searches with di-leptonic top and anti-top events as the dominant background.
\end{abstract}

\thispagestyle{empty}
\newpage

\setcounter{page}{1}

\baselineskip18pt

\vspace{-3cm}

%%%%%%%%%%%%%%%%%%%%%%%%%%%%%%%%%%%%%%%%%%%%%%%%%%%
\section{Introduction}
\label{sec:introdction}

The hierarchy problem of the standard model (SM) has become more prominent with the discovery of a light Higgs boson at the Large Hadron Collider (LHC)~\cite{Aad:2012tfa, Chatrchyan:2012ufa}. On the other hand, no new physics which could address the hierarchy problem has shown up in the searches at the LHC so far. Supersymmetry (SUSY) is widely viewed as the most promising solution to the hierarchy problem. However, the canonical SUSY scenarios are already highly constrained by the null results of many SUSY searches at the LHC. In particular, the gluino and the first two generation squarks are constrained to be heavier than $\sim 1$~TeV if they have the standard $R$-parity conserving decays~\cite{Chatrchyan:2013sza, ATLAS-CONF-2012-109,Sekmen:2011cz,Arbey:2011un,Strubig:2012qd,CahillRowley:2012rv, CahillRowley:2012kx}. The third generation squarks, on the other hand, have weaker bounds. Not only the mass reaches are lower than those of the first two generation squarks, but also in the low mass region there is still plenty of parameter space not being experimentally excluded, due to the more complicated decay patterns and SM backgrounds. Since they have the largest direct couplings to the Higgs fields, they are the most relevant superpartners for the hierarchy problem. Naturalness arguments strongly prefer them to be light, independent of any particular SUSY scenario. Therefore, they have been the focus of intensive experimental and phenomenological investigations nowadays.

Studies of stop searches have been performed by many groups in various (fully hadronic, single-lepton, di-lepton) channels in recent years \cite{Kitano:2006gv, Barbieri:2009ev, Brust:2011tb, Papucci:2011wy, Kane:2011zd, Essig:2011qg, Berger:2011af, Carena:2008mj, Bornhauser:2010mw, Kats:2011it, He:2011tp, Drees:2012dd, Han:2008gy, Plehn:2010st, Plehn:2011tf, Bai:2012gs, Plehn:2012pr, Alves:2012ft, Shelton:2008nq, Han:2012fw, Kaplan:2012gd, Brust:2012uf, Cao:2012rz, Chen:2012uw, Evans:2012bf, Bi:2012jv, Franceschini:2012za, Graesser:2012qy, Dutta:2013sta, Ian2013, Asano:2010ut, Espinosa:2012in, Krizka:2012ah, Kilic:2012kw, Dutta:2012kx, Chakraborty:2013moa, Choudhury:2012tc, Choudhury:2012kn}.  The LHC has done dedicated stop searches \cite{ATLAS-CONF-2012-166, ATLAS-CONF-2012-167, CMS-PAS-SUS-12-023, ATLAS-CONF-2013-037, ATLAS-CONF-2013-024, ATLAS-CONF-2013-001}, but so far has not found any evidence for the stop particles.  As the fine tuning in the Higgs sector increases with the stop mass, an intriguing possibility is that the stop is not very heavy, but the mass spectrum of the stop and other lighter superpartners is somewhat compressed and the event signals from stop decay do not look very different from the SM backgrounds.  In that case, simple variables often used in SUSY searches, such as $\met$ and the effective mass, are usually not very helpful in discriminating signals from backgrounds. It is therefore desirable to look for more sophisticated variables which may have more discriminating power.  In this paper, we focus on such scenarios, where the mass difference between the stop and neutralino is close to or smaller than the top mass.   In this case, the decay of the stop to a bottom quark and a chargino can be dominant, since the decay to a top and a neutralino is closed or phase-space suppressed.  We only consider model-independent direct stop pair production by assuming all other colored superpartners are heavy. The signal event topology that we study is then a symmetric decay chain in which a pair of stops both decay through a chargino, which decays into a stable neutralino plus a $W$ gauge boson.  Furthermore, we focus on the di-leptonic channel, in which both $W$-bosons (from the decay of charginos) decay to a lepton-neutrino pair.

The ATLAS collaboration at the LHC has done such a search with $13 \mbox{~fb}^{-1}$ of data at 8 TeV~\cite{ATLAS-CONF-2012-167} (and an update with $21 \mbox{~fb}^{-1}$~\cite{ATLAS-CONF-2013-024} just appeared while we finalize our paper).  In that analysis, an $M_{T2}$\cite{Lester:1999tx, Barr:2003rg} variable defined with the leptons, is the primary variable throughout the analysis.  The $M_{T2}$ of leptons has an end point around $M_W$ for the $t\bar{t}$ background, and in general has a strong correlation with the $p_T$'s of the leptons.  Hence, it is very useful when the signal has harder leptons than the background does.  However, depending on the mass spectrum, the signal does not necessarily have harder leptons, especially for the compressed scenario.  It is important to explore different regions of parameter space, including those regions that would produce soft leptons (but maybe hard $b$-jets), which would in general require different strategies and kinematic variables.  In this paper we choose a few representative points in the parameter space with different characteristics and try to identify the best strategy for each case.  The rest of the paper is organized as follows.  In Sec.~\ref{sec:basic}, we present the sample signal spectra, and describe the details of event simulations and basic selections.  In Sec.~\ref{sec:newvariables}, we define many kinematic variables which can be constructed from the kinematic information of the events. The distributions of the signals and the background in these variables are shown so that we can see which variables can be useful for different mass spectra. In Sec.~\ref{sec:performance}, we perform a detailed study of these kinematic variables one by one and also in combinations to compare their performances in various scenarios. We identify the best variables and strategies for each of the sample spectra. The conclusions are drawn in Sec.~\ref{sec:conclusions}. In Appendix~\ref{app:topmix}, we show how the results change when some nonzero branching fraction of the stop decay through the top quark and the neutralino is included. Some details of the definition and computation of a new set of variables are described in Appendix~\ref{app:dd}.

%%%%%%%%%%%%%%%%%%%%%%%%%%%%%%%%%%%%%%%%%%%%%%%%%%%
\section{Sample Spectra, Event Simulations and Basic Event Selections}
\label{sec:basic}
\begin{figure}[th!]
\centering
\includegraphics[width=5cm]{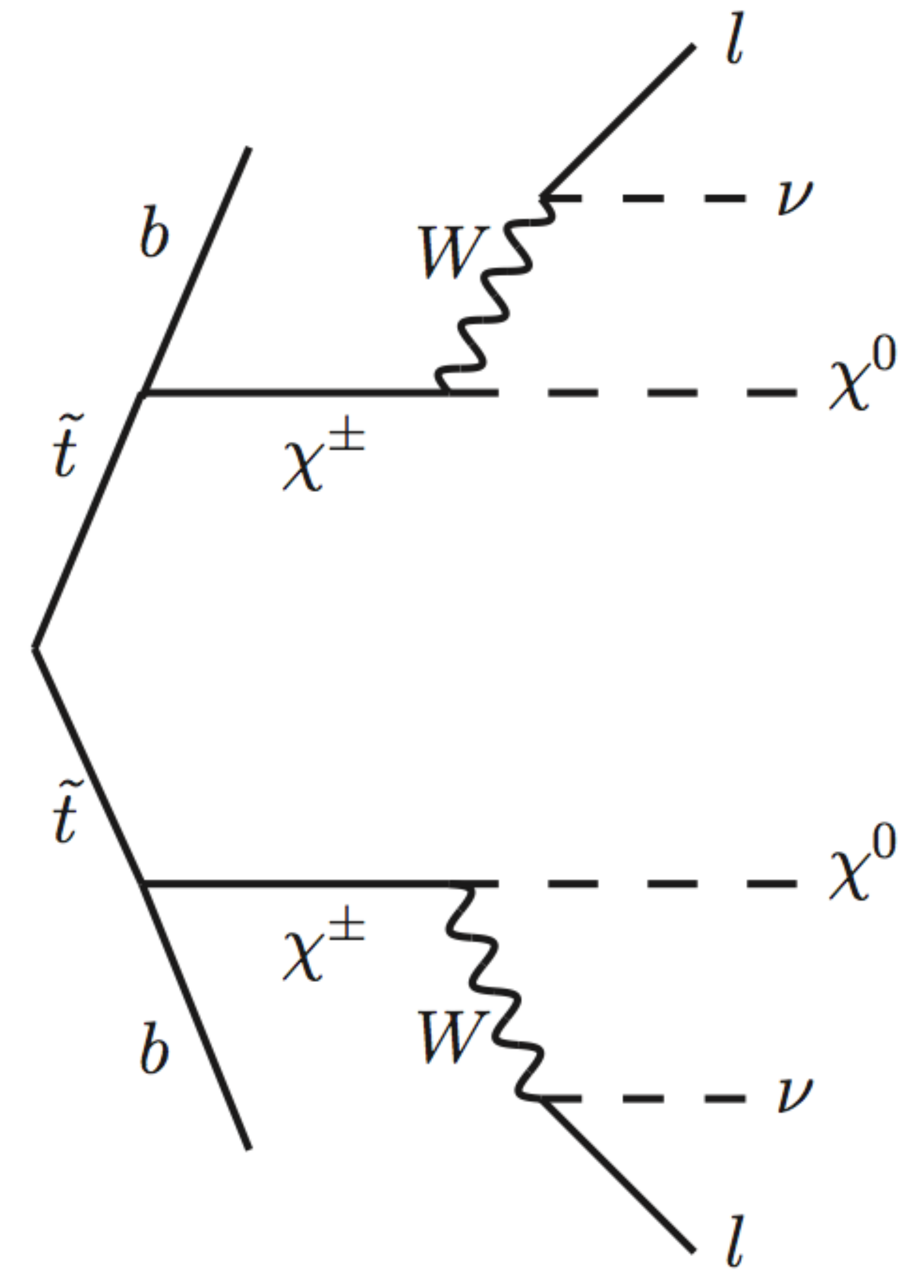}
\hspace{2cm}
\includegraphics[width=5.5cm]{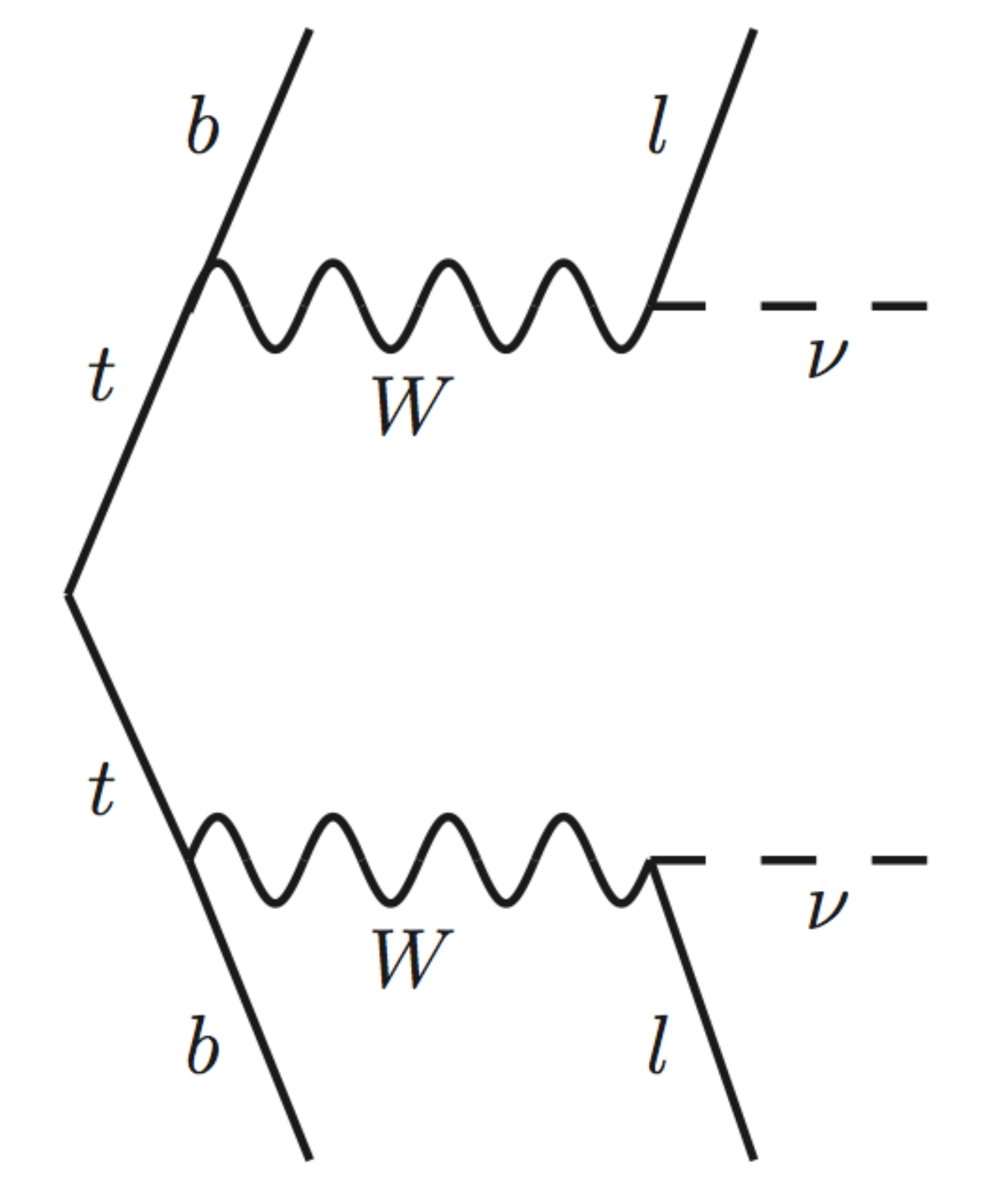}
\caption{Left: Feynman diagrams of stop pair production with both stops decaying through a chargino and with 2 leptons final states.  Right: Feynman diagrams of $t\bar{t}$ pair production with two leptons final states, which is the main background after some basic cuts.}
\label{fig:feyn}
\end{figure}

%%%%%%%%%%%%%%%%%%%%%%%%%%%%%%%%
The stop particles can be produced in pairs at the LHC via the QCD interaction. If $R$-parity is conserved, they will decay into other lighter $R$-parity odd superpartners. In this paper, we assume that some of the neutralinos and charginos are lighter than the stop and focus on the challenging scenario with a relatively small mass difference between the lightest stop and the lightest neutralino: $m_W < m_{\tilde{t}_1} - m_{\tilde{\chi}^0_1} \lesssim m_t$. If the mass difference is large, the visible decay products will be very energetic and easily distinguishable from the SM backgrounds, then the search is mostly limited by the production rate. On the other hand, if the spectrum is highly degenerate, the decay products may be too soft to pass the triggers or experimental cuts. One has to rely on the mono-jet search to place bounds in that case. Assuming a light chargino sitting in between the stop and neutralino masses, $m_{\tilde{\chi}^0_1} < m_{\tilde{\chi}^\pm_1} < m_{\tilde{t}_1}$, the dominant decay channel of the stop is likely to be $\tilde{t}_1 \rightarrow b \tilde{\chi}^\pm_1$ with the chargino $\tilde{\chi}^\pm_1$ decaying into $\tilde{\chi}^0_1$ plus an on-shell or off-shell $W$ gauge boson. Concentrating on the case of leptonic decays for both $W$ bosons, we have the signal final state to be $2b+2\ell+E_T^{\rm miss}$. 
The detailed Feynman diagram for the di-leptonic stop signal is shown in the left panel of Fig.~\ref{fig:feyn}. 
The dominant background for the stop search in this channel is the $t\bar{t}$ production with leptonic decays in both decay chains after requiring a minimum of the missing traverse energy, as shown in the latest ATLAS analysis~\cite{ATLAS-CONF-2012-167,ATLAS-CONF-2013-024} (though the ATLAS analysis used a cut on a $M_{T2}$ variable instead of missing transverse energy to suppress other backgrounds). The corresponding Feynman diagram is shown in the right panel of Fig.~\ref{fig:feyn}. In our study, we focus on the $t\bar{t}$ background, and look for suitable kinematic variables which can effectively separate the signal from the $t\bar{t}$ background in order to improve the stop search in this channel. 

The $b$-jet and lepton momenta in the final state follow from the two mass differences: $m_{\tilde{t}_1} - m_{\tilde{\chi}^\pm_1}$ and $m_{\tilde{\chi}^\pm_1}-m_{\tilde{\chi}^0_1}$. The equivalent mass differences for the $t\bar t$ background are $m_t - M_W$ and $M_W - m_\nu$ which are fixed by the top and $W$ masses. As a consequence, we anticipate that for different spectra one may use different kinetic variables to improve the search. To illustrate this point, we study several representative spectra based on whether the $b$-jet and the lepton are harder or softer in the final state. We choose two different mass gaps between the stop and the neutralino. The first one has $300-120=180$~GeV, which is close to the top quark mass. The other one has $250-120=130$~GeV with a smaller mass gap than that of the top background and generically softer leptons and $b$-jets. For each fixed mass gap between the stop and the neutralino, we study three different cases with the chargino mass close to the stop mass, close to the neutralino mass, or at a point with similar mass differences as the $W$ gauge boson mass between the top and the neutrino masses. The six sets of masses are shown in Table~\ref{tab:spectrum}, where we label the six different spectra from S1 to S6. We also highlight the characteristic features of the $b$-jet and lepton momenta by comparing them to the momenta in the $t\bar t$ background. We choose the mass differences such that even for softer $b$-jets or leptons, a significant fraction of them can still pass the cuts and register in the signal events, otherwise alternative search strategies will be needed. All of these six spectra are not ruled out by the ATLAS search at 8~TeV with 13~fb$^{-1}$, although the spectrum S3 may have been excluded by the 21~fb$^{-1}$ update~\cite{ATLAS-CONF-2013-024}. We want to emphasize that the current search strategies at ATLAS~\cite{ATLAS-CONF-2012-167} using the $M_{T2}$ variable constructed from the harder leptons will only be sensitive to the S3-like spectra. For other type of spectra, different variables are generally required to distinguish the signal from the background. We will come back to this point in Section~\ref{sec:performance}.

\begin{table}[!tb]
   \centering
  \renewcommand{\arraystretch}{1.3}
   \begin{tabular}{|c|c|c|c|c|c|}
     \hline \hline
     & \hspace{0.2cm}  $m_{\tilde{t}_1}$ (GeV)\hspace{0.2cm}   & $m_{\tilde{\chi}^\pm_1}$ (GeV) & $m_{\tilde{\chi}^0_1}$ (GeV) & $b$-jets   & leptons  \\ \hline \hline
     S1 & 300 & 160 & 120 & harder  & softer \\ \hline
     S2 & 300 & 200 & 120 & comparable & comparable \\ \hline
     S3 & 300 & 230 & 120 & softer & harder \\ \hline
     S4 & 250 & 160 & 120 & comparable & softer \\ \hline
     S5 & 250 & 180 & 120 & softer & softer \\ \hline
     S6 & 250 & 200 & 120 & softer & comparable \\ \hline   \hline
   \end{tabular}
   \caption{Six representative signal spectra according to their similarities to the $t\bar t$ background. In the last two columns, the labels, ``softer", ``harder" and ``comparable", mean the comparison of the $b$-jet or lepton momenta between the signal and the $t\bar t$ background. } 
  \label{tab:spectrum}
\end{table}

Our detailed collider studies are based on the 8~TeV LHC with 22~fb$^{-1}$, which is roughly the total integrated luminosity collected by either ATLAS or CMS for the 8~TeV run. Signal and background events are generated using \texttt{MadGraph5}~\cite{Alwall:2011uj}, and showered in
\texttt{PYTHIA}~\cite{Sjostrand:2006za}.  We use \texttt{PGS}~\cite{PGS} to perform the fast detector simulation with modified $b$-tagging efficiencies that roughly match the latest ATLAS $b$-tagging efficiency~\cite{ATLAS:2012aoa}. For signal events, we do not include the $\tau$ leptons from $W$ gauge boson decays, but they are kept in the background events. This is because this type of background may become important once the background with direct electrons and muons from $W$ decays are sufficiently suppressed. The signal production cross section is normalized to be the value calculated at NLO+NLL~\cite{Beenakker:2010nq}.\footnote{\url{https://twiki.cern.ch/twiki/bin/view/LHCPhysics/SUSYCrossSections8TeVstopsbottom}}  We assume the stop decays to a bottom quark and a chargino with $100\%$ branching ratio.  In Appendix~\ref{app:topmix} we also study the case for which the branching ratio of the stop decaying to a top quark and a neutralino is nonzero.  The $t\bar{t}$ production cross section is normalized to be $238^{+22}_{-24}$~pb, calculated at NLO+NNLL~\cite{Aliev:2010zk} (and used by \cite{ATLAS-CONF-2012-167}). We generated $10^5$ events for each signal spectrum, and $5\times 10^5$ events for the $t\bar{t}$ di-leptonic background (which is close to the total number of the corresponding background events of 22~fb$^{-1}$ integrated luminosity).

For the basic cuts on the objects, we closely follow the ATLAS analysis~\cite{ATLAS-CONF-2012-167}:  electrons are required to have $p_T>20$ GeV and $|\eta|<2.47$; muons are required to have $p_T>10$ GeV and $|\eta|<2.4$; jets are required to have $p_T>20$ GeV and $|\eta|<2.5$.  Any jet within $\Delta R\equiv \sqrt{\Delta\phi^2+\Delta\eta^2}=0.2$ of an electron is discarded.  Any electron or muon within $\Delta R=0.4$ of any remaining jet is discarded.  We require exactly two opposite-charge leptons (electrons or muons), and at least one lepton with $p_T>25$ GeV.  The invariant mass of the two leptons need to be larger than $20$ GeV.  If the two leptons have the same flavor, the invariant mass is further required to be outside the $71-111$ GeV window to cut out on-shell $Z$ backgrounds.  Two additional variables, $\Delta\Phi_{\rm min}$ and $\Delta\Phi_b$, are defined in Ref.~\cite{ATLAS-CONF-2012-167} as follows:  $\Delta\Phi_{\rm min}$ is the azimuthal angle difference between the ${\bf p}^{\rm miss}_T$ vector and the closest jet three-momentum;  $\Delta\Phi_b$ is the azimuthal angle difference between the ${\bf p}^{\rm miss}_T$ vector and the vector ${\bf p}^{\ell\ell}_b = {\bf p}^{\rm miss}_T +{\bf p}^{\ell_1}_b +{\bf p}^{\ell_2}_b$.  We require $\Delta\Phi_{\rm min}>1$ and $\Delta\Phi_b<1.5$.  The ATLAS analysis in Ref.~\cite{ATLAS-CONF-2012-167} has focused on the energetic leptons from the signals and did not impose any cut on the $b$-tagged jets. In contrast, we will treat the $b$-jets on the same footing as the lepton, because they are required to construct other variables for different spectra. Specifically, we require at least two $b$-tagged jets. If there are more than two $b$-jets, which rarely happens, we simply select the leading two $b$-jets for our analysis. 
The $b$-jet requirement also reduces backgrounds which contain no $b$-jet.
Different from the ATLAS analysis in Ref.~\cite{ATLAS-CONF-2012-167} where a lower $M_{T2}$ cut was used, we impose a cut on the $E^{\rm miss}_T$ to be above 40~GeV to suppress additional SM backgrounds like $Z+$jets, for which a small $E^{\rm miss}_T$ can be induced from the mis-measurement of jets. 

After the basic cuts, we present the numbers of signal and background events as well as the initial significances $s/\sqrt{b}$ in Table~\ref{tab:numbers}. To obtain the normalized events at 22~fb$^{-1}$, we adopt the following signal production cross sections at the 8~TeV LHC: 2.00~pb for $m_{\tilde{t}_1}=300$~GeV and 5.60~pb for $m_{\tilde{t}_1}=250$~GeV. As can be seen from Table~\ref{tab:numbers}, none of the six spectra can be excluded at 90\% C.L. at the 8~TeV LHC with just the basic cuts. 
\begin{table}[th!]
   \centering
     \renewcommand{\arraystretch}{1.3}
   \begin{tabular}{|c|c|c|c|}
     \hline \hline
      & \# of simulated events   & \# of events at 22~fb$^{-1}$ & $s/\sqrt{b}$  \\ \hline \hline
     S1 & 2718 & 54 & 0.48 \\ \hline
     S2 & 5580 & 112 & 0.99 \\ \hline
     S3 & 4920 & 98 & 0.88 \\ \hline
     S4 & 1538 & 86 & 0.77 \\ \hline
     S5 & 2446 & 137 & 1.22 \\ \hline
     S6 & 2379 & 133 & 1.19 \\ \hline   \hline
     $t\bar t$   & 11344 & 12602 &   \\ \hline   \hline
   \end{tabular}
   \caption{The signal and background events after the basic cuts at the 8~TeV LHC. The second column corresponds to the numbers of signal and background events normalized to the 22~fb$^{-1}$ luminosity. The signal significances are shown in the last column. Signal spectra, S1-S6, have beed described in Table~\ref{tab:spectrum}. } 
  \label{tab:numbers}
\end{table}

%%%%%%%%%%%%%%%%%%%%%%%%%%%%%%%%%%%%%%%%%%%%%%%
\section{New  Kinematic Variables}
\label{sec:newvariables}
%%%%%%%%%%%%%%%%%%%%
In this section, we discuss some traditional and new kinematic variables which may help with the stop search in the chargino decay channel, and examine the distributions of the signal and $t\bar{t}$ background events in these variables. Because of the diversities of the signal spectra, we do not anticipate that any single variable can cover all spectra. Therefore, we explore as many variables as possible and identify which set of variables will work best for each spectrum. 
We categorize the variables that we consider into three different classes: basic variables, $M_{T2}$-based variables, and new ``compatible-masses'' variables which can be viewed as generalizations of $M_{T2}$ variables but using all available kinematic constraints. We will describe each class in sequence. There are also variables which depend on different spins of the top and the stop. A spin-correlation variable, the azimuthal angle between the two leptons, is also investigated at the end of this Section.

%%%%%%%%%%%%%%%%%%%%%%%%%%%%%%%%%%%%%%%%%%%%%%%
\subsection{Basic Variables}
\label{sec:basicvariables}
\begin{figure}[ht!]
\centering
\includegraphics[width=7cm]{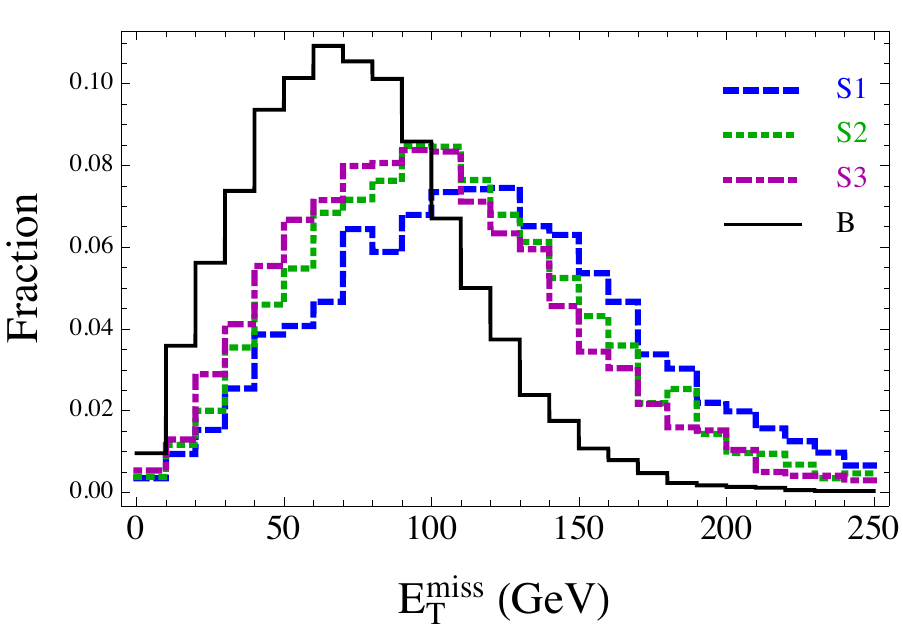}
\hspace{1cm}
\includegraphics[width=7cm]{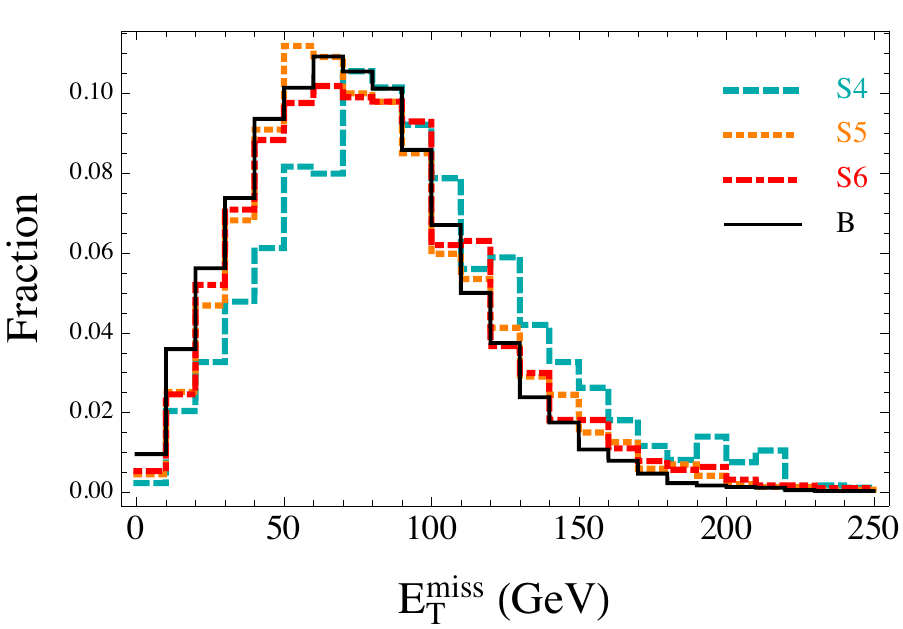} \\
\caption{$\met$ distributions of the six different signals and the $t\bar t$ background.}
\label{fig:met}
\end{figure}

%%%%%%%%%%%%%%%%%%%%
As we saw in Table~\ref{tab:spectrum} that different spectra have different $p_T$'s for $b$-jets and charged-leptons. The $p_T$'s of visible particles can certainly be used to reduce the $t\bar t$ background. The missing transverse energy in a signal event comes from both the neutralinos and neutrinos, while in the background event it only comes from the neutrinos. Consequently, we may expect some differences in $E_T^{\rm miss}$ between the signals and the background. We show their distributions for the $t\bar t$ background and signals for all six spectra in Fig.~\ref{fig:met}, where the horizontal axis starts from zero $E_T^{\rm miss}$. 
From Fig.~\ref{fig:met}, we can see that for S1, S2 and S3, the signal events turn out to have harder distributions of $E_T^{\rm miss}$ than the background and $E_T^{\rm miss}$ could be useful to improve the search. For S4, S5 and S6 with a smaller mass difference of $m_{\tilde{t}_1}-m_{\tilde{\chi}^0_1}$, the difference is much smaller, so cutting on $E_T^{\rm miss}$ is not likely to improve the signal significance. 

\begin{figure}[ht!]
\centering
\includegraphics[width=7cm]{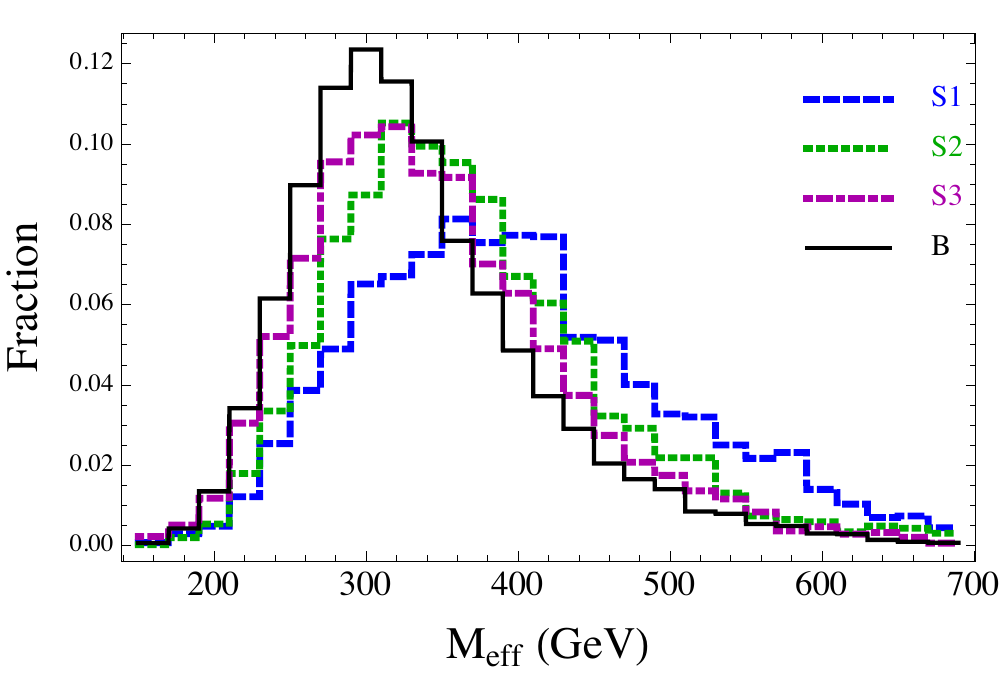}
\hspace{1cm}
\includegraphics[width=7cm]{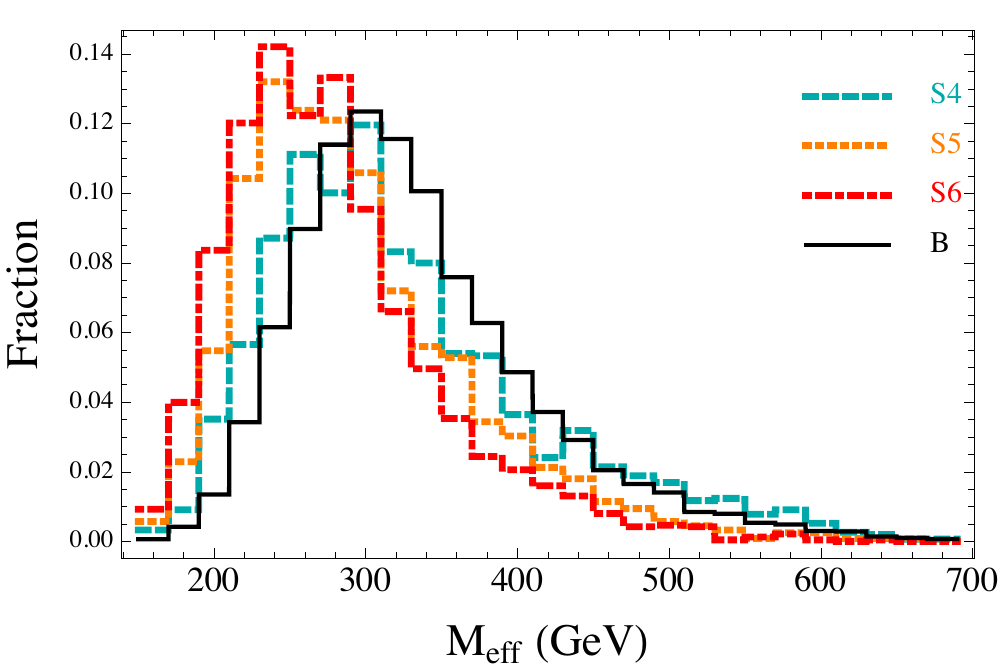} \\
\caption{$M_{\rm eff}$ distributions of the six different signals and the $t\bar t$ background.}
\label{fig:meff}
\end{figure}

Another commonly used kinematic variable is the so-called effective mass variable, defined as
\beqa
M_{\rm eff} \equiv E_T^{\rm miss} + \sum_{i=b_1, b_2, \ell_1, \ell_2}|\vec{p}^i_T| \, ,
\eeqa
which can capture the generic center-of-mass energy of each event and is usually useful for searching for new physics with heavy particles. We show the signal and background distributions in $M_{\rm eff}$ in Fig.~\ref{fig:meff}.
We see that S1, S2, and S3 have somewhat harder distributions than the background. On the other hand, S4, S5, and S6 on average have smaller $M_{\rm eff}$ than the background due to the small splitting between the stop and neutralino masses.

%%%%%%%%%%%%%%%%%%%%%%%%%%%%%%%%%%%%%%%%%%%%%%%
\subsection{$M_{T2}$-based Variables}
\label{sec:mt2variables}
%%%%%%%%%%%%%%%%%%%%
A second class of variables is based on the $M_{T2}$ variable discussed in the literature~\cite{Lester:1999tx, Barr:2003rg, Cheng:2008hk, Barr:2009jv, Burns:2008va, Barr:2011xt, Mahbubani:2012kx}. $M_{T2}$ uses the visible particle momenta and ${\bf p}^{\rm miss}_T$ to find an optimized transverse mass for both decay chains. Specifically for the final state of our interests, one can define the $M_{T2}^\ell$ variable using the two lepton momenta and ${\bf p}^{\rm miss}_T$ as
\beqa
M_{T2}^\ell = \mbox{min}\left\{ \bigcup_{ {\bf p}_1 + {\bf p}_2 = {\bf p}^{\rm miss}_T }  \mbox{max} 
\left[ m_T({\bf p}^{\ell_1}_T, {\bf p}_1), m_T({\bf p}^{\ell_2}_T, {\bf p}_2) \right]   \right\} \,,
\label{eq:MT2l}
\eeqa
where the mass of missing particle is set to be zero, as would be in a $t\bar t$ background event.
\begin{figure}[ht!]
\centering
\includegraphics[width=6.5cm]{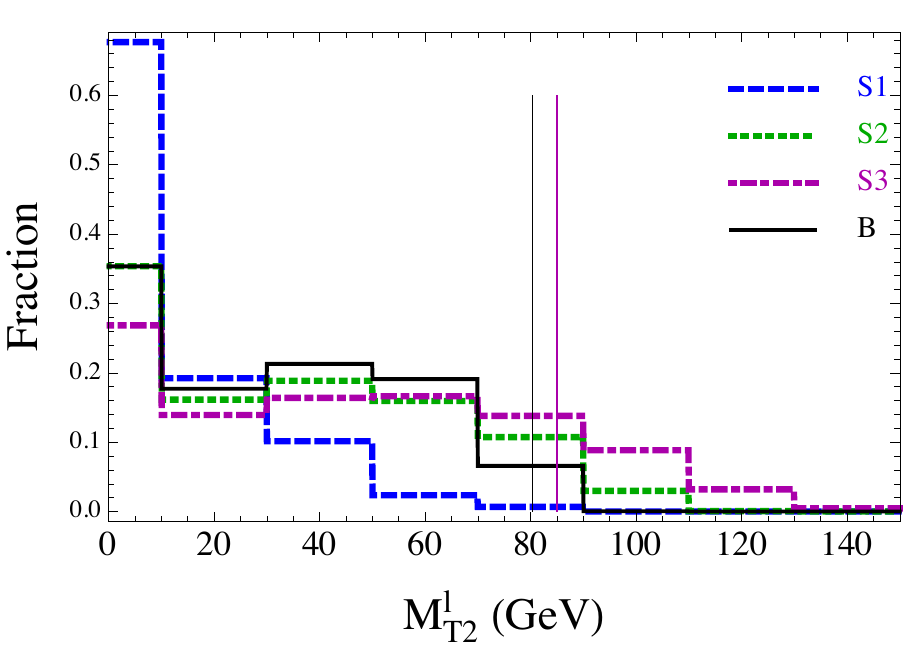}
\hspace{1cm}
\includegraphics[width=6.5cm]{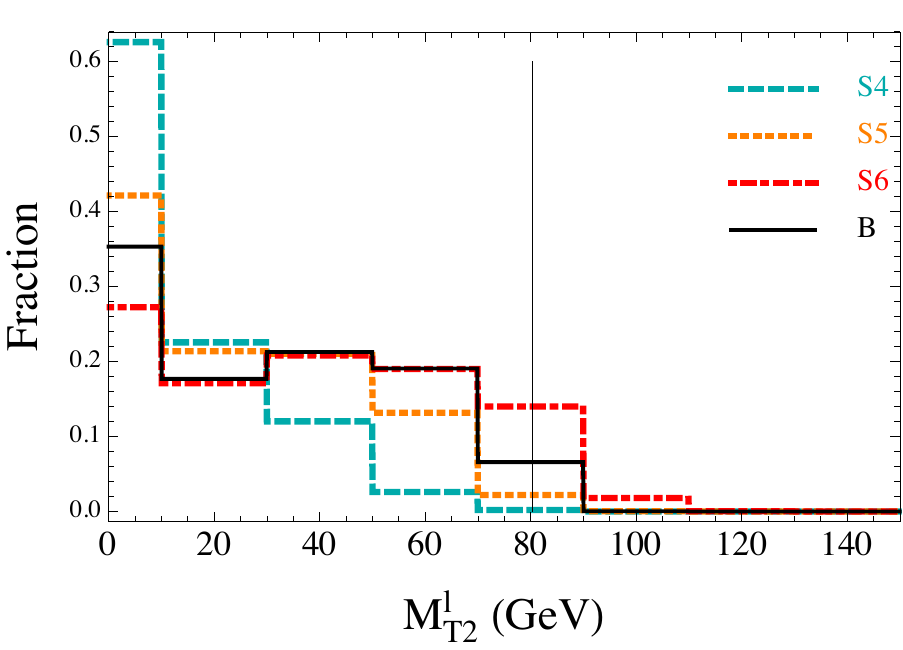} \\
%
%\vspace{0.5cm}
%
%\includegraphics[width=6.5cm]{mt2lbS123.pdf}
%\hspace{1cm}
%\includegraphics[width=6.5cm]{mt2lbS456.pdf}
%
\caption{$M^\ell_{T2}$ distributions for the six signals and the $t\bar{t}$ background (B). The vertical black line indicates the $W$ gauge boson mass. The vertical purple line shows the suggested lower cut to increase the S3 signal significance.  As shown in Ref.~\cite{Lester:2011nj}, $M_{T2}^\ell$ will vanish if ${\bf p}^{\rm miss}_T$ lies in between the two vectors ${\bf p}^\ell_1$ and ${\bf p}^\ell_2$ in the transverse plane. This explains the accumulation of evens at zero. 
}
\label{fig:mT2l}
\end{figure}
The signal and background event distributions are shown in Fig.~\ref{fig:mT2l}.
Comparing the signal and background behaviors in terms of $M_{T2}^\ell$, we can see that S1, S4 and S5 have smaller values. This is essentially due to their softer lepton momenta. As a result, imposing a lower limit cut on $M_{T2}^\ell$, like what has been done in the ATLAS search~\cite{ATLAS-CONF-2012-167}, can only decrease the signal significance. On the other hand, the spectrum S3 does have a harder lepton and a cut of requiring $M_{T2}^\ell$ above a certain value can increase the signal significance. This is indeed the case. We found that imposing $M_{T2}^\ell > 85$~GeV, the signal significance $s/\sqrt{b}$ of S3 increases from 0.88 to 2.94. This increase is very significant and is because the $t\bar t$ background has $M_{T2}^\ell$ bounded from above by the $W$ gauge boson mass (denoted by the vertical black lines in Fig.~\ref{fig:mT2l}).

\begin{figure}[ht!]
\centering
\includegraphics[width=6.5cm]{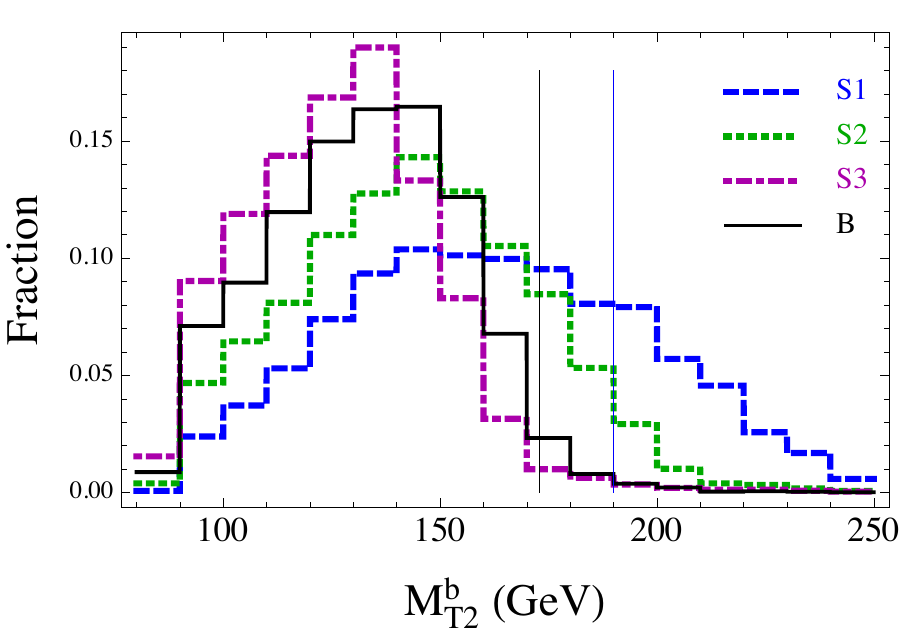}
\hspace{1cm}
\includegraphics[width=6.5cm]{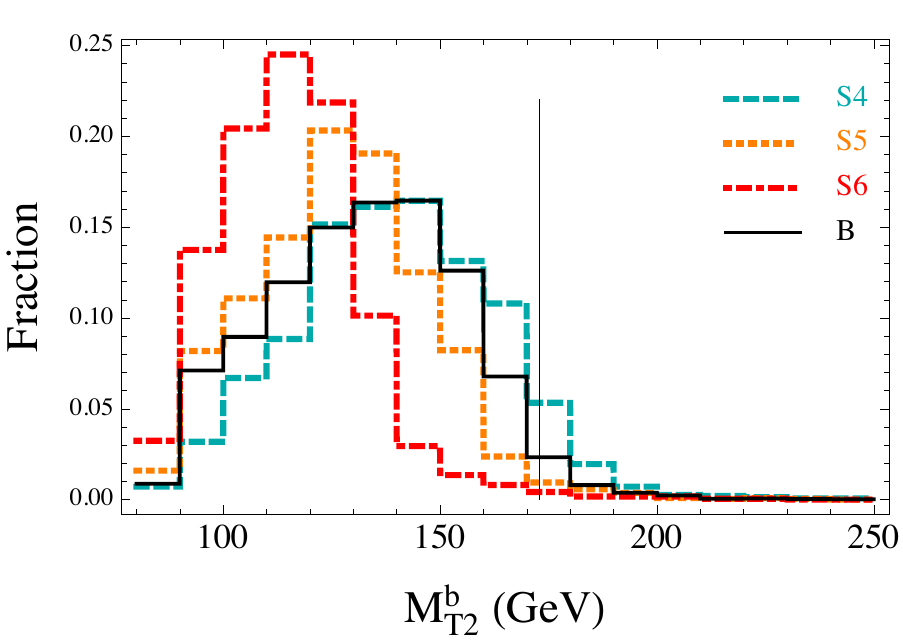}
\caption{$M^b_{T2}$ distributions for the six signals and the $t\bar{t}$ background (B). The vertical black line indicates the top quark mass. The vertical blue line shows the suggested lower cut to increase the S1 signal significance.}
\label{fig:mT2b}
\end{figure}

 A similar variable $M_{T2}^b$ can be constructed from replacing ${\bf p}^{\ell_i}_T$ by ${\bf p}^{b_i}_T$ and ${\bf p}^{\rm miss}_T$ by ${\bf p}^{\rm miss}_T+{\bf p}^{\ell_1}_T+{\bf p}^{\ell_2}_T$ in Eq.~(\ref{eq:MT2l}) and assuming the missing particle mass as the $W$ gauge boson mass. (A general discussion of the subsystem $M_{T2}$ can be found in Ref.~\cite{Burns:2008va}.) We show the $M^b_{T2}$ distributions for signals and $t\bar{t}$ background in Fig~\ref{fig:mT2b}. For the $t\bar t$ background, the $M^b_{T2}$ distribution has an end-point at the top quark mass, as indicated by the vertical black line. The relative signal and background distributions follow from the hardness of the $b$-jet. Since the S1 signal has the hardest $b$-jet, its $M^b_{T2}$ extends to a larger value. Imposing a lower limit cut on $M^b_{T2}$ can therefore substantially increase the signal significance. For example, a cut of $M^b_{T2}>190$~GeV, denoted by the vertical blue line in Fig~\ref{fig:mT2b} can provide an improvement of $s/\sqrt{b}$ from $0.48$ to $1.27$.

\begin{figure}[ht!]
\centering
\includegraphics[width=6.5cm]{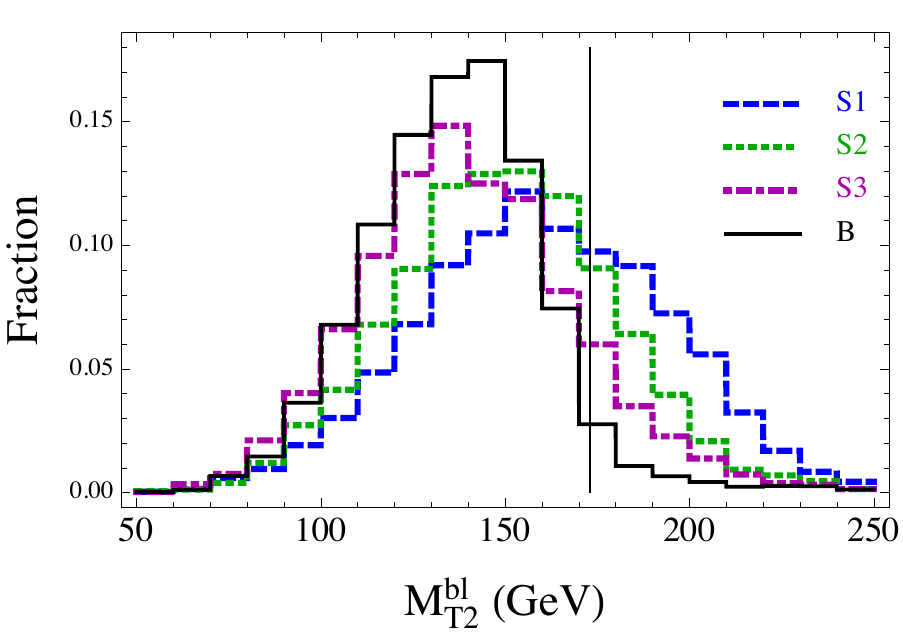}
\hspace{1cm}
\includegraphics[width=6.5cm]{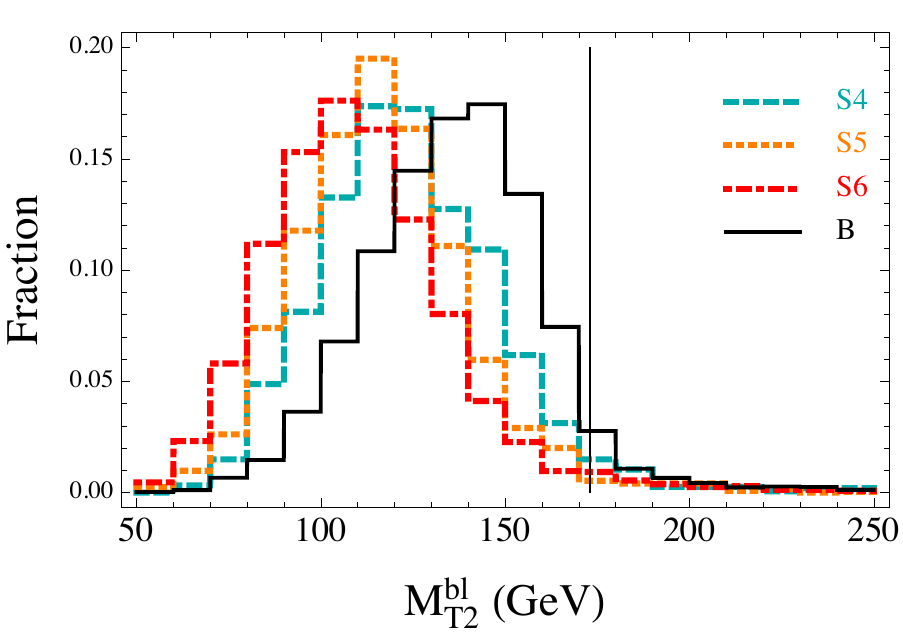}
\caption{$M^{b\ell}_{T2}$ distributions for the six signals and the $t\bar{t}$ background (B). The vertical black line indicates the top quark mass.}
\label{fig:mT2bl}
\end{figure}

One additional $M_{T2}^{b\ell}$ variable can be defined by treating the summation of one $b$-jet momentum and one charged-lepton momentum together as a single effective visible particle and the missing particle in each decay chain as massless. There is a combinatorial ambiguity for this variable on how to choose the two possible pairs. In our analysis, we choose the pairing with a smaller value of the larger invariant mass of the $b$-jet momentum and the lepton momentum between the two pairs for each possible paring. This choice is made based on the fact that for the correct combination the $b$-jet and the lepton come from the same mother particle decay, so their invariant mass should be limited by the mother particle mass. The signal and background distributions are shown in Fig.~\ref{fig:mT2bl}. For the S1, S2, S3 distributions, one can see that the hardness of $b$-jets, leptons and $E_T^{\rm miss}$ can make the signal $M_{T2}^{b\ell}$ distribution extend beyond that of the $t \bar t$ background. On the contrary, the S4, S5 and S6 signal distributions have their majority of $M_{T2}^{b\ell}$ below the top quark mass. So, to increase the signal significance for these spectra, one may want to impose an upper limit cut on $M_{T2}^{b\ell}$ instead.
%

%%%%%%%%%%%%%%%%%%%%%%%%%%%%%%%%%%%%%%%%%%%%%%%
\subsection{Compatible-masses Variables}
\label{sec:fullvariables}
%%%%%%%%%%%%%%%%%%%%
Each of the $M_{T2}$ variables discussed above only uses a part of the full kinematic information of each event. For the signals and background in our study, the two decay chains are symmetric. One can in principle use the three equal on-shell-mass constraints in the two-step decays at the same time to further distinguish signals from background~\cite{Cheng:2007xv}. For example, we can concentrate on the $t\bar t$ background and define variables to fully utilize the top quark, $W$ gauge boson, and neutrino mass constraints. A simple counting of the number of unknowns and the number of equations indicates that the unknown neutrino momenta can be obtained by solving the coupled equations up to a four-fold ambiguity~\cite{Bai:2008sk}. 
As the $M_{T2}$ can be interpreted as the minimal mother particle mass compatible with the kinematics of an event for an assumed event topology and the daughter particle mass~\cite{Cheng:2008hk}, we can consider the two mass-squared differences in the two-step decays that are compatible with all visible particle momenta and the equal-on-shell-mass constraints. They define an allowed region in the two-dimensional parameter space for each event. There is more than one way to extract useful variables from such distributions. We choose to define $\Delta_1$ and $\Delta_2$ as the compatible mass-squared differences which minimize their sum (i.e., the mass-squared difference between the first particle and the last particle in the two-step decay chain). Similar to the $M_{T2}^{b\ell}$ variable, there is a combinatorial issue for this variable in choosing the two possible pairs of $b$-jet and lepton momenta. In our analysis, we choose the pair with a smaller value of the larger invariant mass of the $b$-jet momentum and the lepton momentum as in the case for the $M_{T2}^{b\ell}$. The detailed definition and computation of these two variables are described in Appendix~\ref{app:dd}. For the $t \bar t$ background, $\Delta_1$ and $\Delta_2$ provide an estimate of $M_W^2 - m_\nu^2$ and $m_t^2 - M_W^2$ respectively. It turns out, even after the detector smearing and the ambiguity of the event reconstruction, $\Delta_1$ and $\Delta_2$ have a scattered distribution near the true mass-squared differences.

\begin{figure}[ht!]
\centering
\includegraphics[width=7cm]{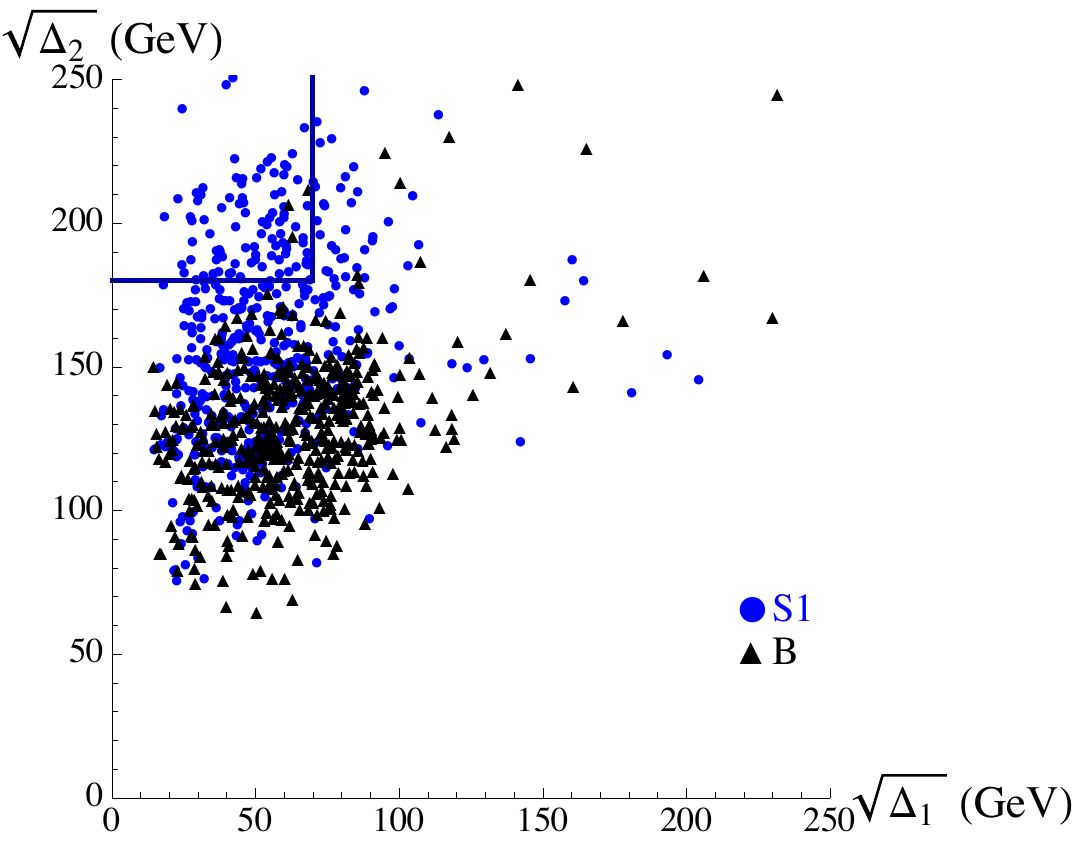}
\hspace{0.4cm}
\includegraphics[width=7cm]{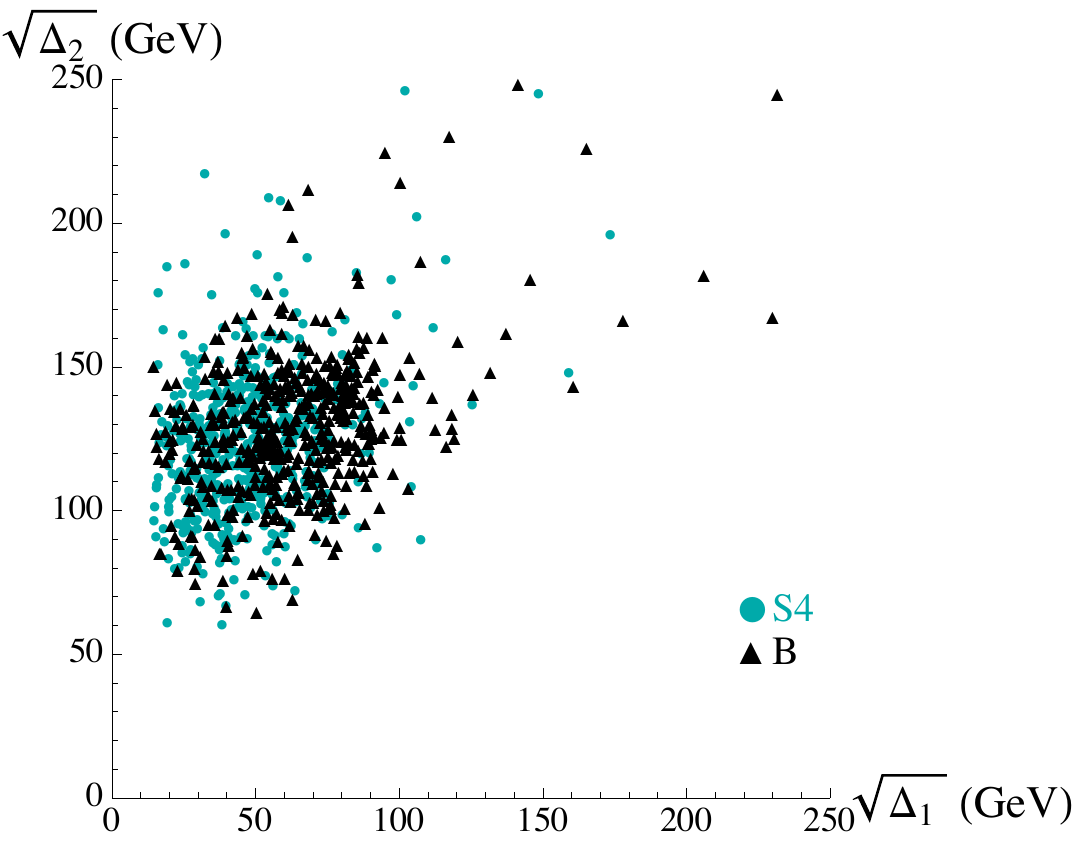}  \vspace{0.4cm} \\
\includegraphics[width=7cm]{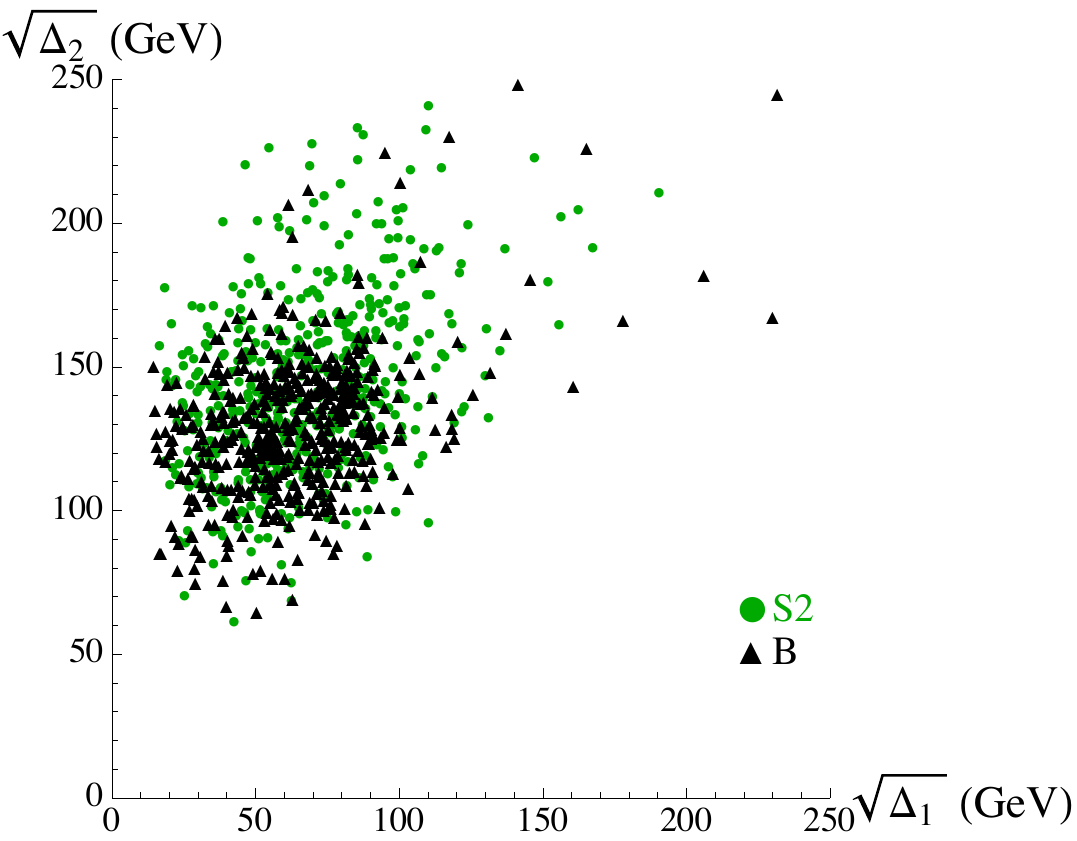}
\hspace{0.4cm}
\includegraphics[width=7cm]{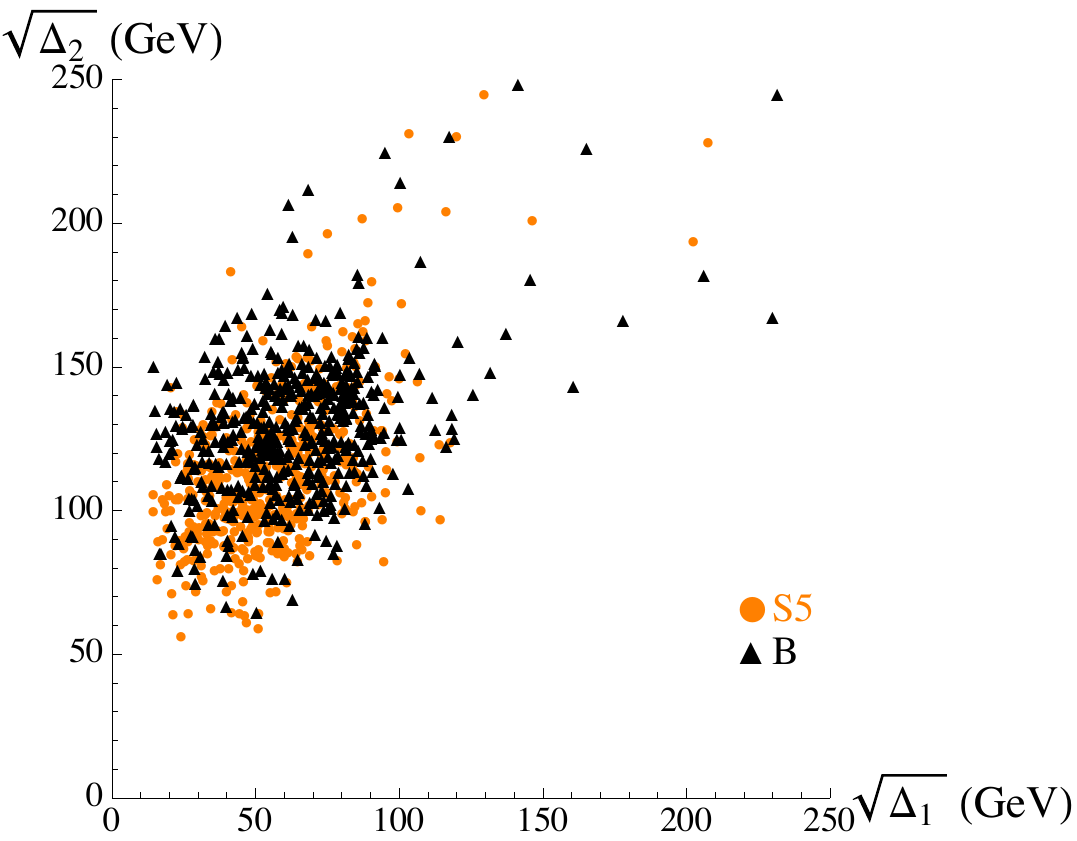}  \vspace{0.4cm} \\
\includegraphics[width=7cm]{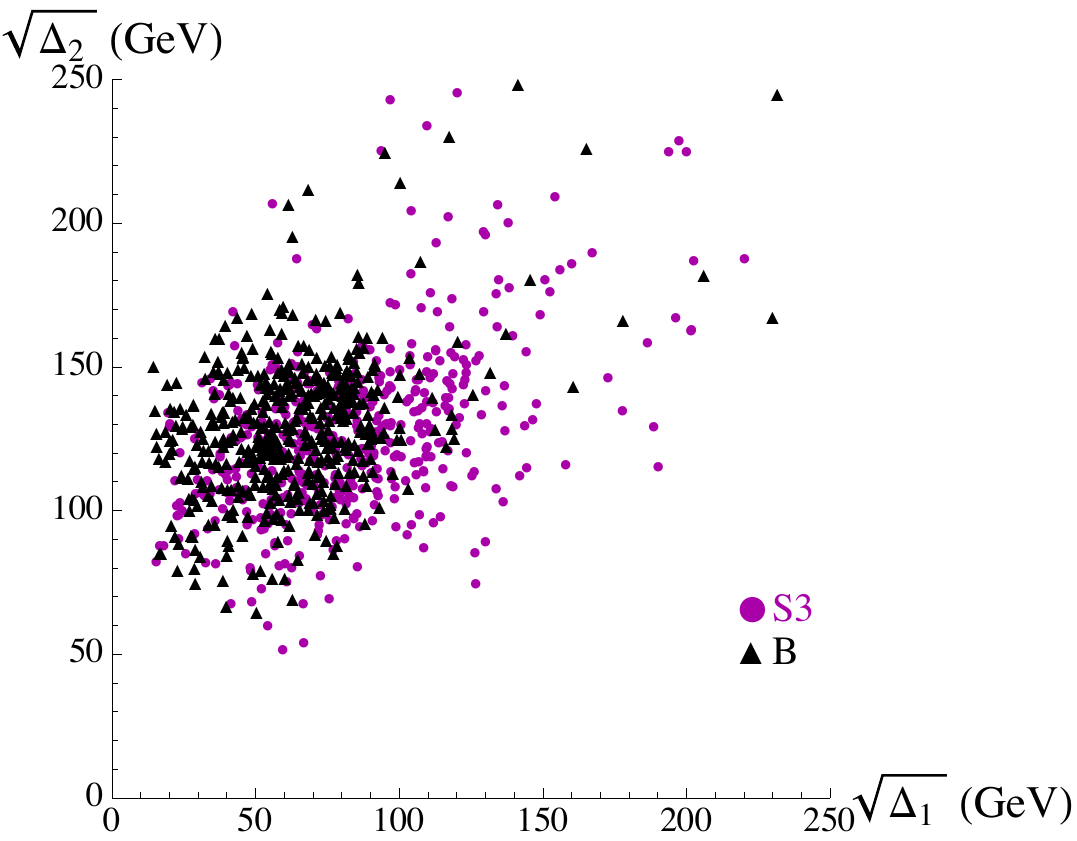} 
\hspace{0.4cm}
\includegraphics[width=7cm]{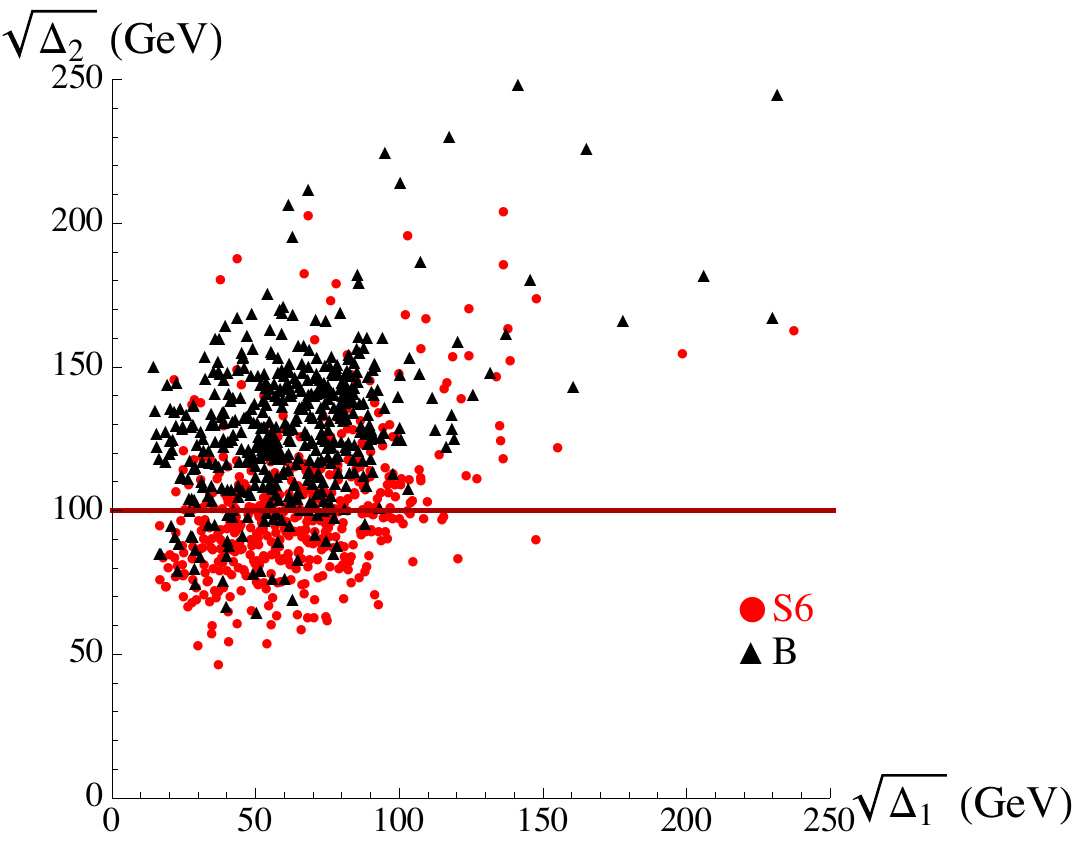}
\caption{Scatter plots for S1-6 and $t\bar{t}$ background (B) in the $(\sqrt{\Delta_1}, \sqrt{\Delta_2})$ plane.  500 points are shown for both the signal and the background.}
\label{fig:ddS123456}
\end{figure}

For the signal events, $\Delta_2$ will be close to $m_{\tilde{t}_1}^2 - m_{\tilde{\chi}^\pm_1}^2$, while $\Delta_1$ is closely related to but does not have a simple formula in terms of $m_{\tilde{\chi}^\pm_1}$ and $m_{\tilde{\chi}^0_1}$ because of the two different sources contributing to the total missing transverse energy. As a result, the signal distributions are expected to be more scattered than the background. We show the scatter plots of $\sqrt{\Delta_1}$ and $\sqrt{\Delta_2}$ for the six different signals vs.\ background in Fig.~\ref{fig:ddS123456}. Looking at these scatter plots, we can already see that S1 and S6 have the center of points dramatically different from the $t\bar t$ background. For the signal S1, we found that imposing a cut of $\sqrt{\Delta_1}<70$~GeV and $\sqrt{\Delta_2}>180$~GeV (indicated by the blue vertical and horizontal lines) can increase the significance from $0.48$ to $1.32$. For the signal S6, imposing a cut $\sqrt{\Delta_2} < 100$~GeV can increase the significance from 1.19 to 2.07.

%%%%%%%%%%%%%%%%%%%%%%%%%%%%%%%%%%%%%%%%%%%%%%%
\subsection{A Spin Correlation Variable}
\label{sec:dphi}
%%%%%%%%%%%%%%%%%%%%

\begin{figure}[ht!]
\centering
\includegraphics[width=7cm]{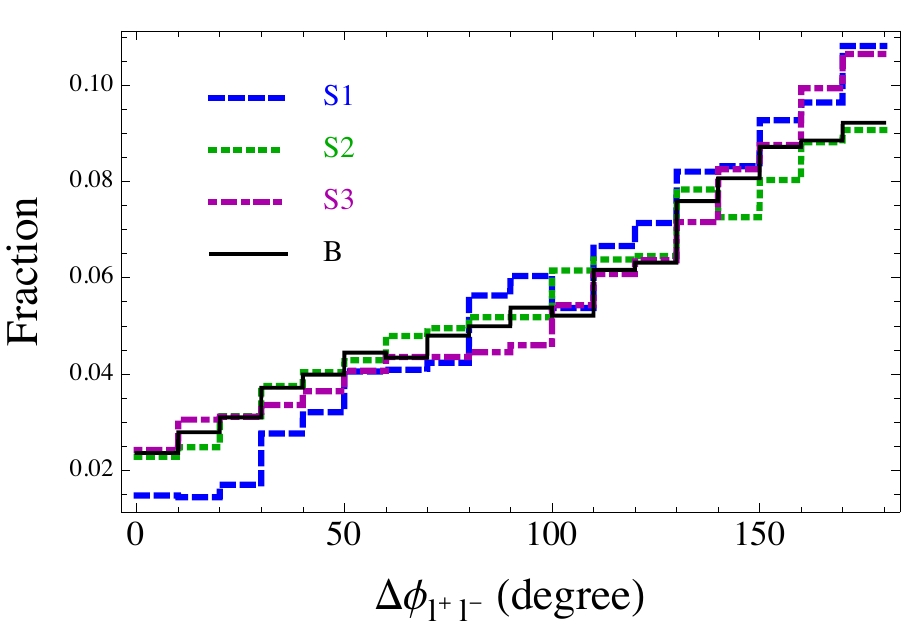}
\hspace{1cm}
\includegraphics[width=7cm]{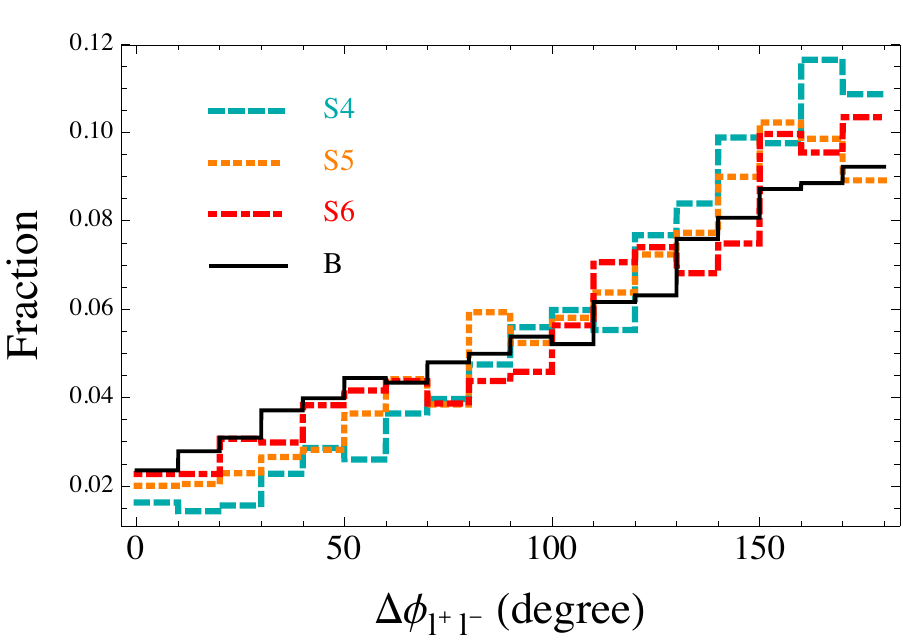} \\
\caption{$\Delta\phi_{\ell^+\ell^-}$ distributions for the six different signals and the $t\bar t$ background.  $\Delta\phi_{l^+l^-}$ is the azimuthal angle difference of the two leptons in the final state.}
\label{fig:dphi}
\end{figure}

In addition to the spectrum difference, one may also explore possible kinematic effects due to different spins of the top and the stop to distinguish signals from the background. Such effects can arise from both production and decays. Top quarks are generally produced with larger rapidity gaps than the stops and Ref.~\cite{Han:2012fw} has exploited this as a possible discriminator of top and stop events. However, it suffers from large systematic uncertainties and the result was not particularly promising. Another effect that might be useful is the spin correlation.  The top quark has spin-1/2 and the top-anti-top pair is produced with correlated polarizations. The leptons in the final states inherit the spin correlation from the top quarks and can be used to constructed variables sensitive to this effect~\cite{Shelton:2008nq, Han:2012fw, Ian2013}. On the other hand, stops are scalars and the decays from the two stops are not correlated at all. In Ref.~\cite{Han:2012fw}, the azimuthal angle difference between the two leptons, $\Delta\phi_{\ell^+\ell^-}$, has been found to be useful for the case with stop decaying to the top quark and neutralino. We apply this variable to our case with the stop decaying to the bottom quark and chargino and plot the distributions of $\Delta\phi_{\ell^+\ell^-}$ for the six different signals and the $t\bar t$ background in Fig.~\ref{fig:dphi}.  For all signals and background, the $\Delta\phi_{\ell^+\ell^-}$ distributions tend to peak at $\pi$, indicating that the two leptons prefer to move in the opposite directions. Although some signals have slightly sharper distributions than the background, we have checked that cutting on $\Delta\phi_{\ell^+\ell^-}$ does not improve the signal significance even when combined with other variables. Therefore, we will not include this spin-correlation variable in the subsequent studies in Sec.~\ref{sec:performance} when we consider combinations of variables.

%

%%%%%%%%%%%%%%%%%%%%%%%%%%%%%%%%%%%%%%%%%%%%%
\section{Variable Performance and Improvements on Different Spectra}
\label{sec:performance}
\begin{figure}[htp!]
\centering
\includegraphics[width=10cm]{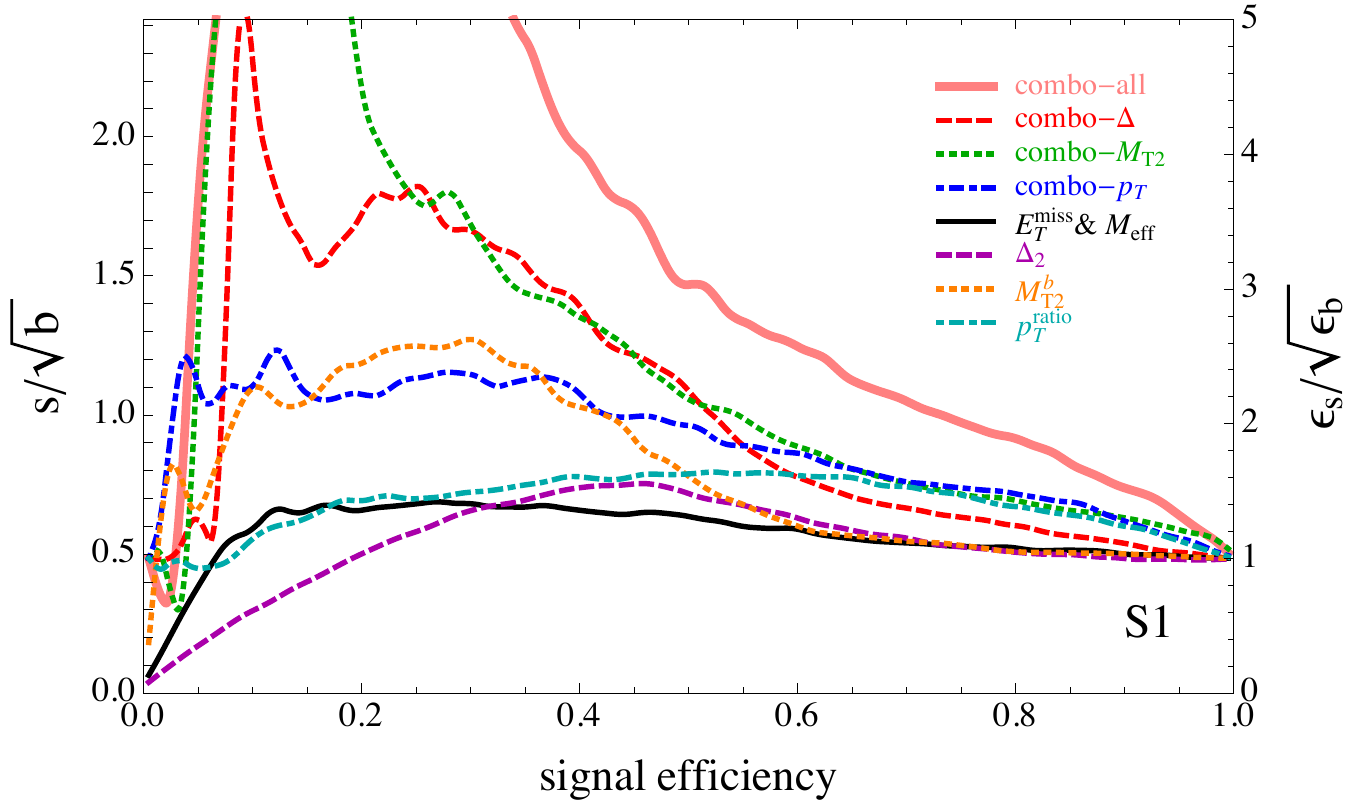} \\
\vspace{0.5cm}
\includegraphics[width=10cm]{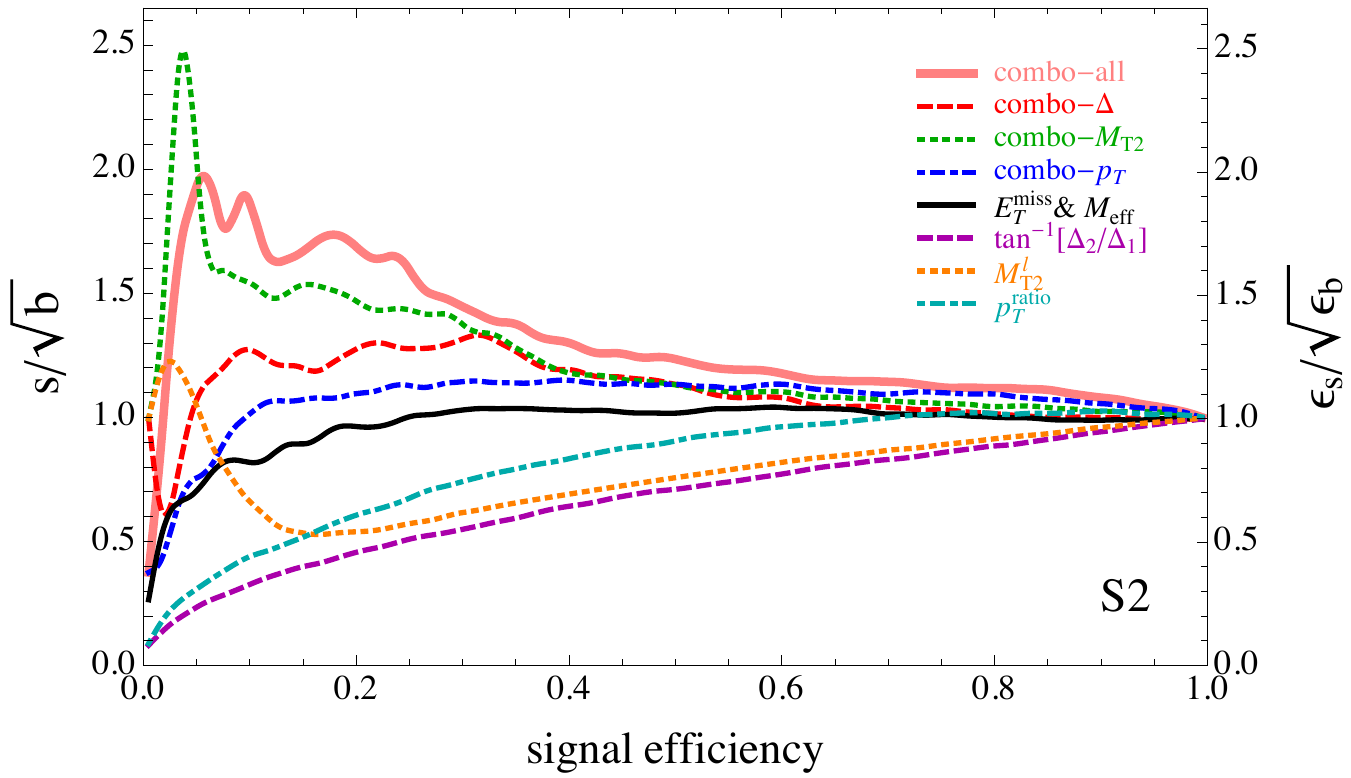} \\
\vspace{0.5cm}
\includegraphics[width=10cm]{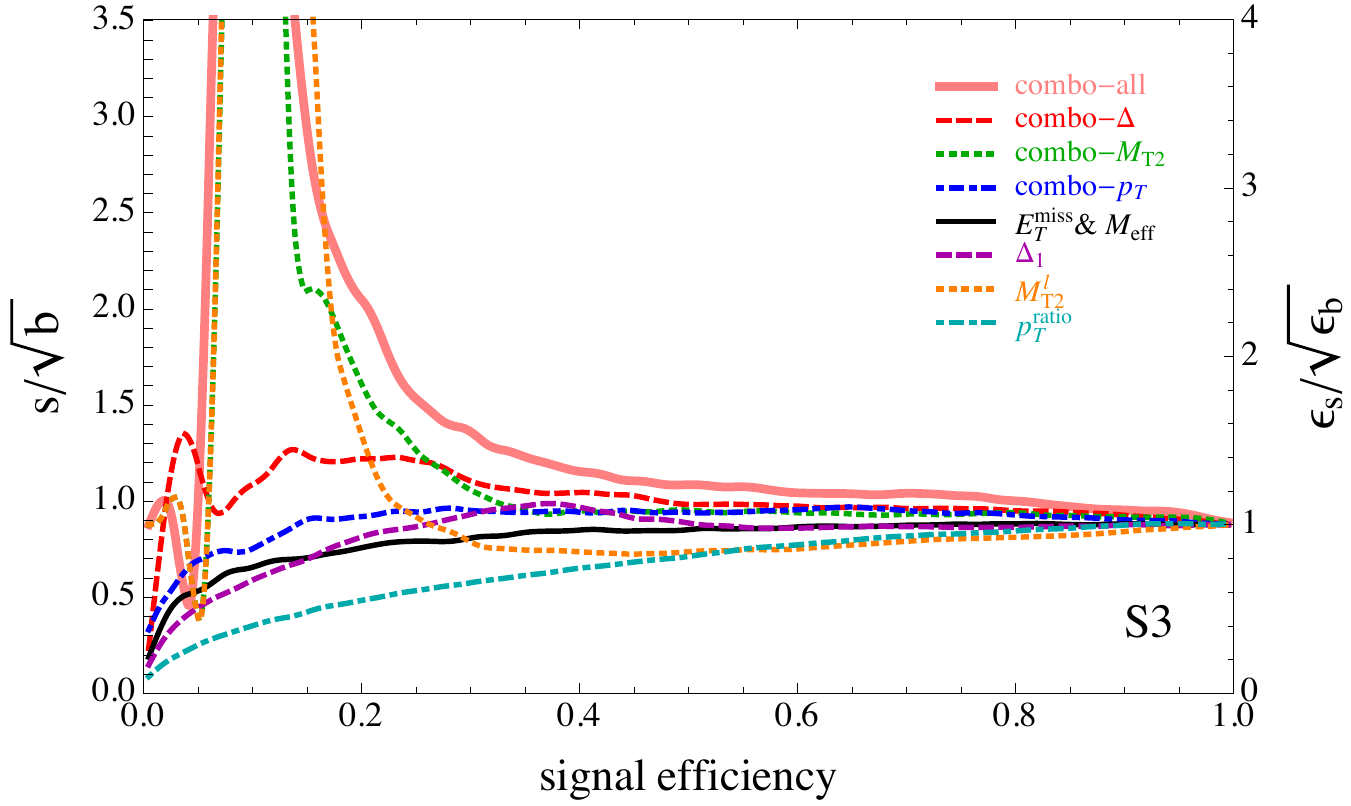}
\caption{The signal significance as a function of the signal efficiency $\epsilon_s$ after the Boost Decision Tree optimization for S1-3. The right vertical label is the efficiency ratio $\epsilon_s/\sqrt{\epsilon_b}$, indicating the relative improvements on top of the basic cuts.}
\label{fig:BDTS123}
\end{figure}

\begin{figure}[htp!]
\centering
\includegraphics[width=10cm]{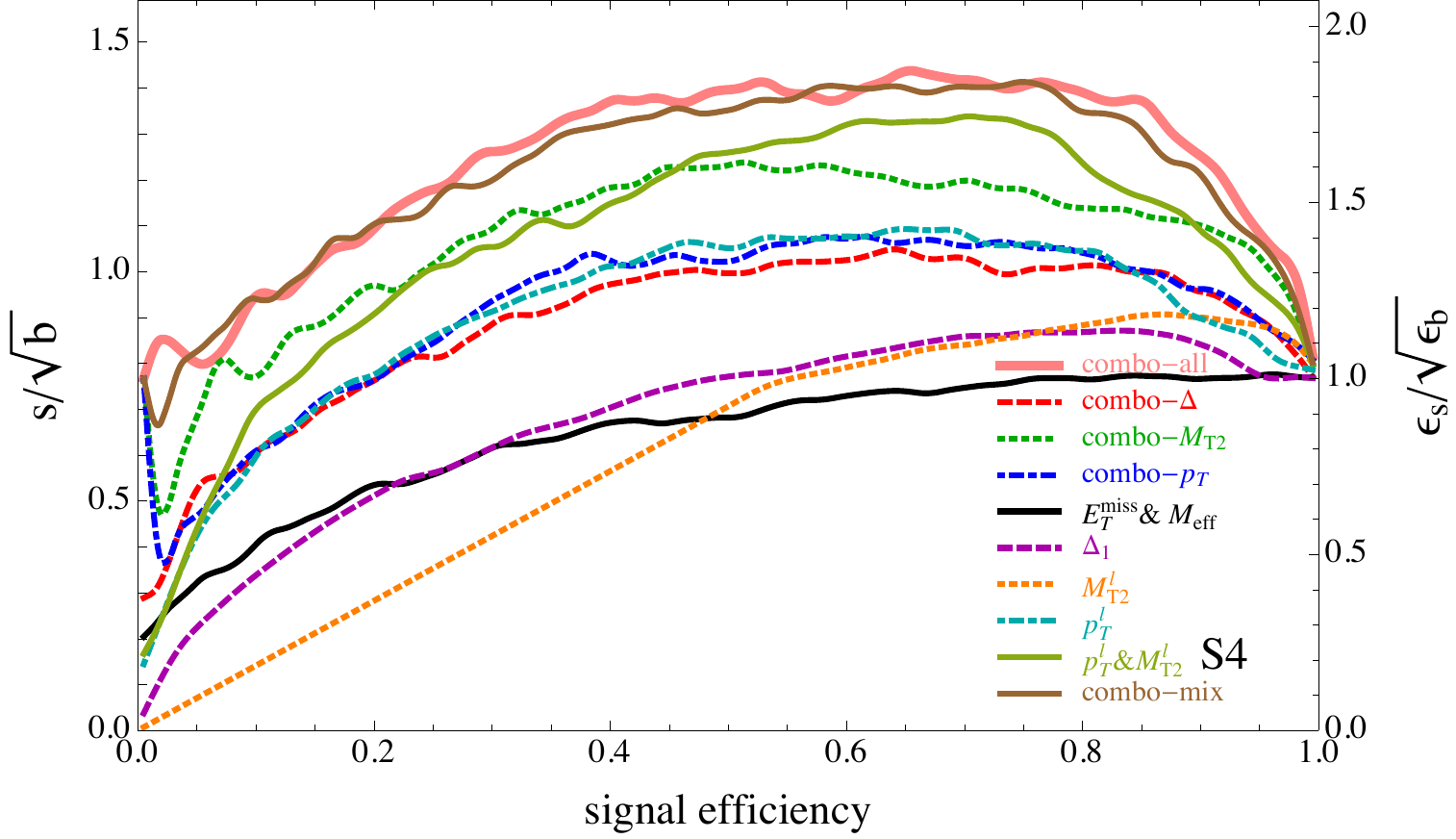} \\
\vspace{0.5cm}
\includegraphics[width=10cm]{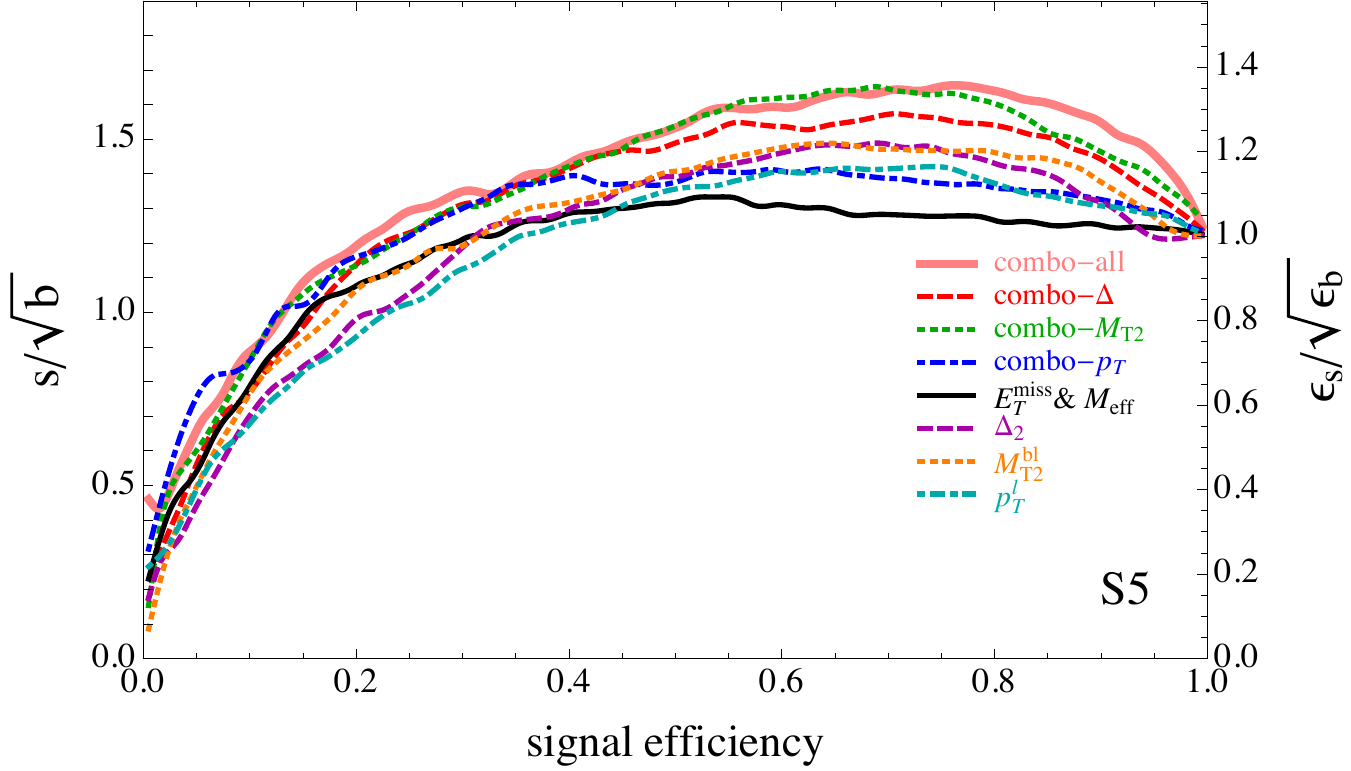} \\
\vspace{0.5cm}
\includegraphics[width=10cm]{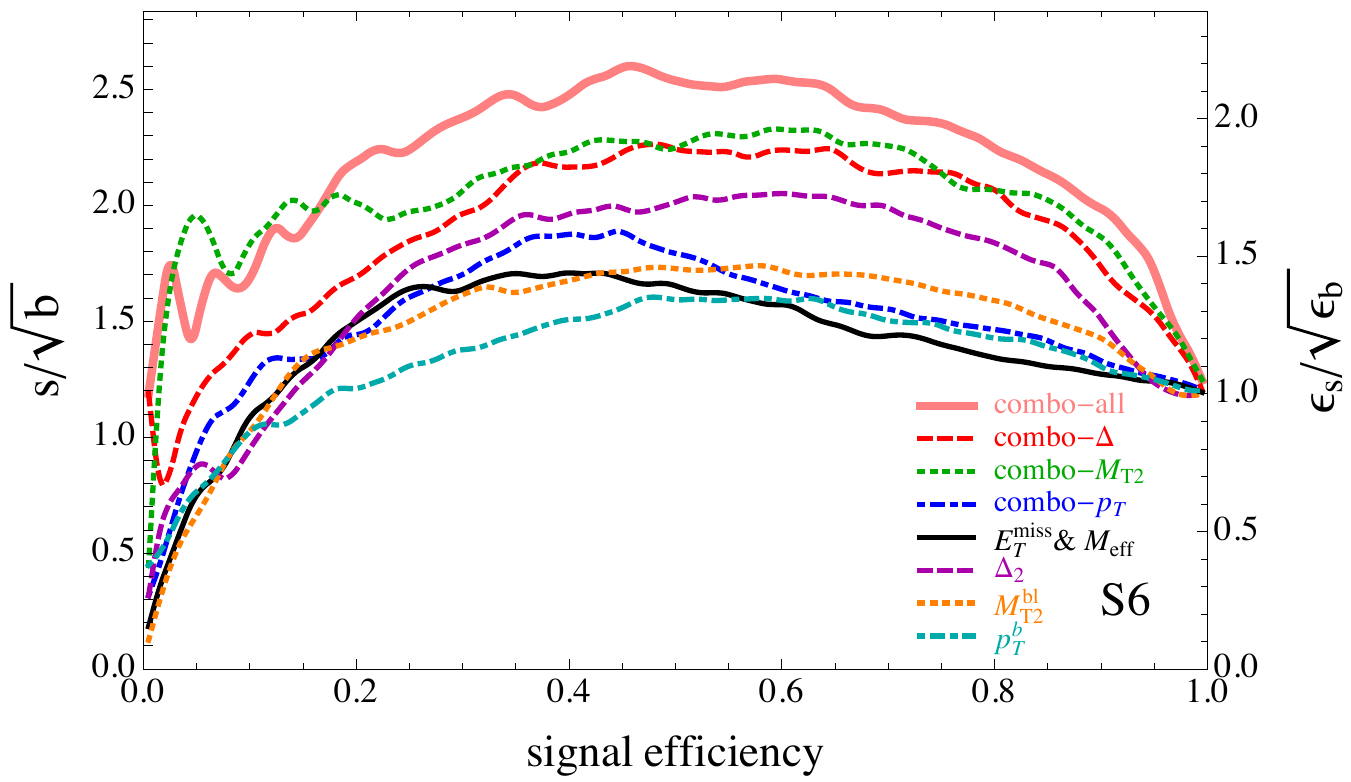}
\caption{Same as Fig.~\ref{fig:BDTS123} but for S4-6. The ``combo-mix" for S4 is a combination of  $\met$, $M_{\rm eff}$, $p^\ell_T$, $M^\ell_{T2}$ and $\Delta_1$. }
\label{fig:BDTS456}
\end{figure}

%%%%%%%%%%%%%%%%%%%%
Having described the individual variables, we now study their performance. For some spectra there is no single variable that works well by itself, so we also combine several variables together and study the improvement on the stop search. Instead of searching for the best cut by hand, we use the Boost Decision Tree (BDT) method~\cite{Roe:2004na} to identify the best variable or the best combination of variables for a given spectrum. Starting from the signal significance after basic cuts in Table~\ref{tab:numbers}, we show the changes of the signal significance $s/\sqrt{b}$ as a function of the signal efficiency for different combinations of variables in Figs.~\ref{fig:BDTS123} and \ref{fig:BDTS456} . 
We have defined different combinations of variables as
\begin{itemize}
  \item combo-all:  the combination of all variables,
  \item combo-$\Delta$:  the combination of $\met$, $M_{\rm eff}$ and $\Delta$ variables,
  \item combo-$M_{T2}$: the combination of $\met$, $M_{\rm eff}$ and $M_{T2}$ variables,
  \item combo-$p_T$: the combination of $\met$, $M_{\rm eff}$ and $p_T$ variables,
  \item $\met \,\&\, M_{\rm eff}$:  the combination of only $\met$ and $M_{\rm eff}$. 
\end{itemize}
In the region with the signal efficiency $\epsilon_s > 0.1$ and the background efficiency $\epsilon_b > 0.01$, where the statistics of events can be trusted, combo-all always provides the best $s/\sqrt{b}$. The right vertical label of these plots in Figs.~\ref{fig:BDTS123} and \ref{fig:BDTS456} is $\epsilon_s/\sqrt{\epsilon_b}$, which directly indicates how much we improve the signal significance from the basic selection of Sec.~\ref{sec:basic}. For S1, S2 and S3, the largest value of $s/\sqrt{b}$ has a small value of the signal efficiency at around 0.1--0.2, which suggests that the optimized cut prefers to reduce both signal and background events. On the contrary, for S4, S5 and S6, the largest value of $s/\sqrt{b}$ has a pretty large value of signal efficiency, which suggests that the best cuts do not occur at the tail of the distributions. 

To identify which variable is the best one and how it behaves in each group, we also show the best variable in each of the $\Delta$, $M_{T2}$ and $p_T$ groups, which consist of 
\begin{itemize}
  \item for the $\Delta$ group: $\Delta_1$, $\Delta_2$, $\sqrt{\Delta_1+\Delta_2}$ and $\tan^{-1}{[\Delta_2/\Delta_1]}$,
  \item for the $M_{T2}$ group: $M^b_{T2}$, $M^\ell_{T2}$ and $M^{b\ell}_{T2}$,
  \item for the $p_T$ group: individual $p_T$'s, $p^b_T\equiv p^{b_1}_T + p^{b_2}_T$,  $p^\ell_T \equiv p^{\ell_1}_T+p^{\ell_2}_T$ and $p^{\rm ratio}_T \equiv p^b_T/p^l_T$.
\end{itemize}
For the moderately compressed spectra considered in this paper, there is no single variable that can reach the same level of improvement as the combination of all variables does. Our results suggest that the best strategy for the experimental analysis in this more difficult region of the model parameter space should at least combine several kinematic variables to obtain the maximal exclusion reach.

\begin{table}[bht!]
   \centering
        \renewcommand{\arraystretch}{1.3}
   \begin{tabular}{|c|c|c|c|c|c|c|c|}
     \hline \hline
& $\epsilon_s$ & $\epsilon_b$ & $\epsilon_s/\sqrt{\epsilon_b}$ & $s$ & $b$ & $s/\sqrt{b}$ & $s/b$     \\ \hline \hline
S1 & $0.405$ & $0.0103$ & $3.98$ & $22.0$ & $130$ & $1.93$ & $0.169$     \\ \hline
S2 & $0.175$ & $0.0101$ & $1.74$ & $19.5$ & $127$ & $1.73$ & $0.154$     \\ \hline
S3 & $0.225$ & $0.0126$ & $2.00$ & $22.1$ & $159$ & $1.76$ & $0.139$     \\ \hline
S4 & $0.655$ & $0.122$ & $1.87$ & $56.4$ & $1540$ & $1.44$ & $0.0366$     \\ \hline
S5 & $0.765$ & $0.318$ & $1.36$ & $105$ & $4009$ & $1.65$ & $0.0261$     \\ \hline
S6 & $0.455$ & $0.0432$ & $2.19$ & $60.6$ & $544$ & $2.60$ & $0.111$     \\ \hline \hline
   \end{tabular}
   \caption{$s/\sqrt{b}$ and $s/b$ after the optimized cuts from BDT at 22~fb$^{-1}$, with the requirement that the signal and the background efficiencies to be $\epsilon_s \geq 0.1$ and $\epsilon_b \geq 0.01$.} 
  \label{tab:bdt1}
\end{table}
\begin{table}[bht!]
   \centering
           \renewcommand{\arraystretch}{1.3}
   \begin{tabular}{|c|c|c|c|c|c|c|c|}
     \hline \hline
& $\epsilon_s$ & $\epsilon_b$ & $\epsilon_s/\sqrt{\epsilon_b}$ & $s$ & $b$ & $s/\sqrt{b}$ & $s/b$     \\ \hline \hline
S1 & $0.355$ & $0.00552$ & $4.78$ & $19.3$ & $69.6$ & $2.31$ & $0.277$     \\ \hline
S2 & $0.175$ & $0.0101$ & $1.74$ & $19.5$ & $127$ & $1.73$ & $0.154$     \\ \hline
S3 & $0.185$ & $0.00549$ & $2.50$ & $18.2$ & $69.2$ & $2.19$ & $0.263$     \\ \hline
S4 & $0.655$ & $0.122$ & $1.87$ & $56.4$ & $1540$ & $1.44$ & $0.0366$     \\ \hline
S5 & $0.765$ & $0.318$ & $1.36$ & $105$ & $4009$ & $1.65$ & $0.0261$     \\ \hline
S6 & $0.455$ & $0.0432$ & $2.19$ & $60.6$ & $544$ & $2.60$ & $0.111$     \\ \hline \hline
   \end{tabular}
   \caption{The same as Table~\ref{tab:bdt1}, but requiring $\epsilon_s \geq 0.1$ and $\epsilon_b \geq 0.005$.} 
  \label{tab:bdt2}
\end{table}

For S1, the variable $M^b_{T2}$ turns out to be the best single variable as we expected.  The combination of $\Delta$ variables and the combination of $M_{T2}$ variables are both quite useful.  The combination of all variables is even better, indicating that the correlation of $\Delta$ variables and $M_{T2}$ variables are not too strong. As far as the performance of a single variable is concerned, $M^b_{T2}$ can be used to improve the search, although its increase on $s/\sqrt{b}$ is mild as can be seen from the dotted orange curve of the first plot in Fig.~\ref{fig:BDTS123}. The improvement from a combination of all variables is very dramatic and can increase $s/\sqrt{b}$ by more than a factor of five. To make sure that the improvement numbers are statistically reliable, we require $\epsilon_s > 0.1$ and $\epsilon_b > 0.01 (0.005)$, which means that there are more than 100(50) background events left, and show the signal significance of the optimized cuts in Table~\ref{tab:bdt1}(\ref{tab:bdt2}). 

For S2, both $b$-jet and lepton momenta from the signal are comparable to the background. For the region that we trust our statistics of simulated events, there is no single variable can substantially increase the significance.  Combo-$M_{T2}$ shows a peak structure at the low signal efficiency region, which is mainly due to the variable $M_{T2}^\ell$.  Combo-$\Delta$ is not that useful and only increases the significance by around 20\%. Again the combination of all variables does show a moderate improvement and can increase the significance by a factor of 1.74 from Table~\ref{tab:bdt1}(\ref{tab:bdt2}). 

For S3, the story is very simple. The charged lepton momenta from signal events are generally harder than the background. Although the basic kinematic variables do not improve the signal significance, the $M_{T2}^\ell$ variable is seen to increase $s/\sqrt{b}$ dramatically. The improvement curve from the combination of all variable follow the curve from $M_{T2}^\ell$ only. The enhancement factor on $s/\sqrt{b}$ can reach 2.0(2.5) from Table~\ref{tab:bdt1}(\ref{tab:bdt2}).

\begin{figure}[htp!]
\centering
\includegraphics[width=6.5cm]{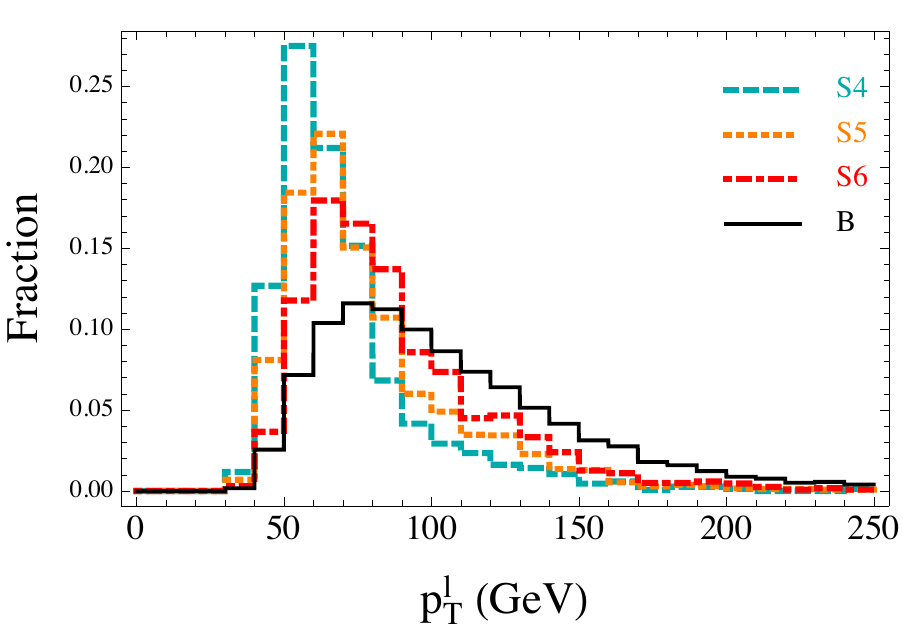}
\hspace{1cm}
\includegraphics[width=6.5cm]{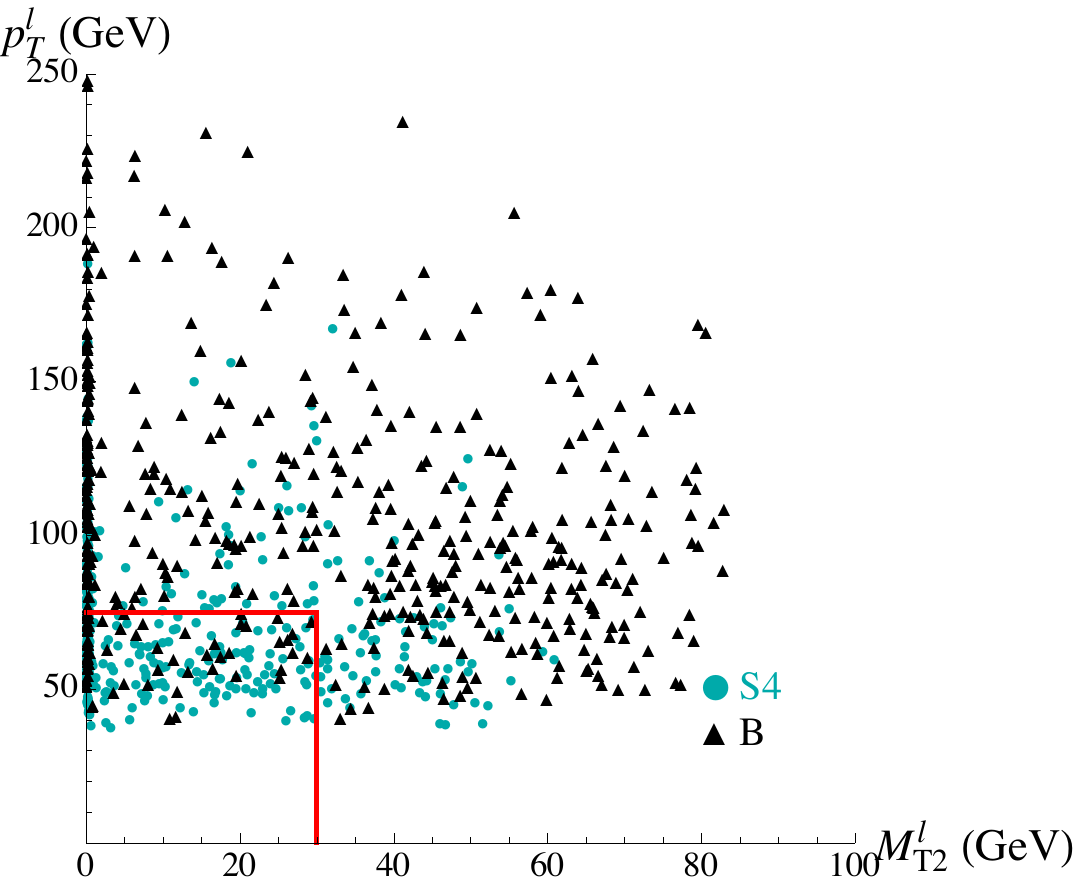}
\caption{Left:  $p^l_T$ (the scalar sum of the $p_T$'s of the two leptons) distributions for S4-6 and the $t\bar{t}$ background (B).  Right: the scatter plot for S4 and B in the $(M^\ell_{T2}, p^\ell_T)$ plane.}
\label{fig:S4}
\end{figure}

For S4, the previously discussed combinations of variables, except the combination of all variables, can hardly improve the signal significance at all. However, the combination of all variables does show an impressive improvement with a factor as large as 1.87 from Table~\ref{tab:bdt1}(\ref{tab:bdt2}). To identify the subset of variables relevant for such an improvement, we have also tried other combinations and found that $p^\ell_T + M^\ell_{T2}$ can do almost as well as the combination of all variables. We show the performance of $p^\ell_T + M^\ell_{T2}$ by the green solid curve in Fig.~\ref{fig:BDTS456}, which almost matches the performance of the combination of all variables. Furthermore, We have found that  the combination of  $\met$, $M_{\rm eff}$, $p^\ell_T$, $M^\ell_{T2}$ and $\Delta_1$ can reach the utmost performance closely. The brown solid curve in Fig.~\ref{fig:BDTS456} shows the performance of this combination, labeled as ``combo-mix." To further understand the behaviors of  $p^\ell_T$ and $M^\ell_{T2}$ for S4, we show the $p^\ell_T$ histogram distribution in the left panel of Fig.~\ref{fig:S4} and the scatter plot of $p^\ell_T$ and $M^\ell_{T2}$ in the right panel.
We can see from the scatter plot that the S4 signal events have a concentration at smaller values of $p^\ell_T$ and $M^\ell_{T2}$. A cut of $M^\ell_{T2}< 30$~GeV and $p^\ell_T < 75$~GeV (indicated by the two red lines in the right panel of Fig.~\ref{fig:S4}) increases $s/\sqrt{b}$ from 0.77 to 1.28.

For the spectrum S5,  there is no variable that provides a substantial improvement of the significance.  Combo-$M_{T2}$ provides a better significance than other combinations, and can do almost as well as the combination of all variables.  The optimized cuts can increase $s/\sqrt{b}$ by a factor of 1.36 and have projected signal significance to be 1.65. As a result, this spectrum could be excluded at 95\% C.L. at 8~TeV with a 22~fb$^{-1}$ luminosity.

For the last spectrum S6, the single variable $\Delta_2$ is better than any other single variables. Combo-$\Delta$ and combo-$M_{T2}$ have similar performances and either one can provide a significant improvement.  The combination of all variables can further improve the signal significance by around 20\% and have the total improvement of 2.19. So, for this type of spectra, the new $\Delta$ variables proposed in this paper (discussed in details in the Appendix~\ref{app:dd}) are highly recommended in the real experimental analysis. 

For the spectra S4-S6, reducing the $t\bar t$ background in general requires upper limit cuts on the variables that we considered because of the softer signal events. One might be concerned about other types of backgrounds not included in this study which may also have softer energies. Most of them have been heavily suppressed by our basic cuts. In addition, the best improvements for S4-S6 occur at relatively high signal efficiencies where the other small backgrounds should not make a large difference. 

%%%%%%%%%%%%%%%%%%%%%%%%%%%%%%%%%%%%%%%%%%%%%%%%%%%
\section{Conclusions}
\label{sec:conclusions}
In this paper, we study the stop search in the chargino decay channel with direct stop production. We focus on the challenging scenario where the spectrum of the superpartners involved in the decay is moderately compressed. The overall transverse momentum of the visible particles (2 leptons + 2 $b$-jets) is similar to or somewhat smaller than that of the $t \bar t$ background, making it difficult to be distinguished from a $t \bar t$ event. However, depending on the intermediate chargino mass, the distribution of the individual lepton or $b$-jet momentum can have different behaviors from the $t \bar t$ background. We studied many kinematic variables, including the simple traditional variables such as $\met$, $M_{\rm eff}$, and individual particle $p_T$'s, the $M_{T2}$ variables, and the new compatible-masses variables ($\Delta_{1,2}$) which use all on-shell conditions of the $t \bar t$ event topology. We found that different variables are useful for different spectra. As a summary, we list the signal spectra, their characteristics of the $b$-jet and lepton momenta and the best variable(s) of each signal spectrum in Table~\ref{tab:best}.
\begin{table}[hbt!]
   \centering
 \renewcommand{\arraystretch}{1.3}
  \begin{tabular}{|c|c|c|c|c|c|c|}
     \hline \hline
     & \hspace{0.2cm}  $m_{\tilde{t}_1}$ (GeV)\hspace{0.2cm}   & $m_{\tilde{\chi}^\pm_1}$ (GeV) & $m_{\tilde{\chi}^0_1}$ (GeV) & $b$-jets   & leptons & best-variables  \\ \hline \hline
     S1 & 300 & 160 & 120 & harder  & softer & $M_{T2}^b$     \\ \hline
     S2 & 300 & 200 & 120 & comparable & comparable & combo-all  \\ \hline
     S3 & 300 & 230 & 120 & softer & harder  & $M_{T2}^\ell$ \\ \hline
     S4 & 250 & 160 & 120 & comparable & softer & $p_T^\ell + M_{T2}^\ell$ \\ \hline
     S5 & 250 & 180 & 120 & softer & softer & combo-all\\ \hline
     S6 & 250 & 200 & 120 & softer & comparable & $\Delta_2$    \\ \hline   \hline
   \end{tabular}   
   \caption{A summary of the best variables for the six different spectra.} 
  \label{tab:best}
\end{table}

A general conclusion from Table~\ref{tab:best} is that different variables should be used for different spectra. Specifically, if either $b$-jets or leptons of the signal are harder than the corresponding ones of the $t\bar t$ background, a single $M_{T2}$ variable can improve the stop search significantly. For example, one could use $M_{T2}^b$ for S1 and $M_{T2}^{\ell}$ for S3. For some spectra like S4, the leptons from the signal are softer than the background, the variables $p_T^\ell$ or $M_{T2}^\ell$ can still be useful to improve the search if one imposes an upper limit cut on these two variables. Similarly for S6, the $b$-jets from the signal are softer. Imposing an upper limit cut on the variable $\Delta_2$ can improve the search. For S2, both $b$-jets and leptons are comparable to those of the background and there is no single variable that works well. For S5, although both b-jets and leptons are softer than the background, the differences are relatively small and there is also no single variable that works well. However, a combination of many variables can still give some improvement for the S2 and S5 spectra.

In our study, we have assumed that all visible particles can pass the basic cuts and be captured for a significant fraction of the signal events. This may no longer be true if the splitting of any two superpartners in the decay chain becomes very small, then the visible particle from the corresponding decay may be too soft to be detected. In that case, one can only rely on the visible particles from the other step of the decay (e.g., $b$-jets if the chargino and neutralino are too degenerate, or leptons if the stop and chargino are too degenerate). The $M_{T2}$ variables might still be useful if the distributions of the signal and background are sufficiently different.  In the limit that all three superpartners are degenerate, few visible particles are hard enough to pass the basic cuts, one has to revert to the mono-jet or mono-photon search with an initial state radiation.

For a final remark, we would like to emphasize that the kinematic variables studied in the paper can also be applied to other new physics searches as long as the dominant background is the $t \bar t$ in the dileptonic channel. For example, one can apply them to the $t^\prime$ search with $t^\prime \rightarrow b + W$ and $W$ decaying leptonically.

%%%%%%%%%%%%%%%%%%%%%%%%%%%%%%%%%%%%%%%%%%%%%%%%%%%
\subsection*{Acknowledgments}
We would like to thank the collaboration of Markus Luty in the early stage of this work, and Jessie Shelton for useful discussion and comments. Y. Bai is supported by startup funds from the UW-Madison. This work of H.-C. Cheng, J. Gallicchio and J. Gu was supported in part by U.S. DOE grant No. DE-FG02-91ER40674.

%%%%%%%%%%%%%%%%%%%%%%%%%%%%%%%%%%%%%%%%%%%%%%%%%%%
\appendix

%%%%%%%%%%%%%%%%%%%%%%%%%%%%%%%%%%%%%%%%%%%%%%%
\section{Including the Decay Channel to Top}
\label{app:topmix}
%%%%%%%%%%%%%%

Throughout our study we have assumed that the stop decays to a bottom quark and a chargino with $100\%$ branching ratio.  This assumption can be a good approximation for the spectra that we study, since the other common decay channel, the stop decaying to a top quark and a neutralino, is highly suppressed by the available phase space. Nevertheless, depending on the couplings and the compositions of these particle, the branching fraction of the decay through the top quark may not be negligible. In this Appendix we study how our results are modified when this decay channel is included.

\begin{figure}[ht!]
\centering
\includegraphics[width=7cm]{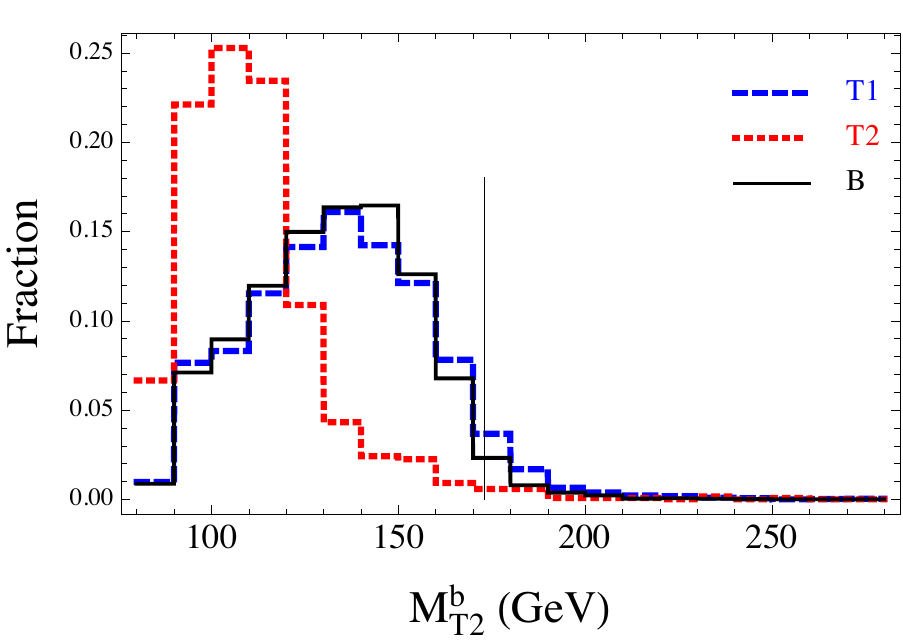} \hspace{0.4cm}
\includegraphics[width=7cm]{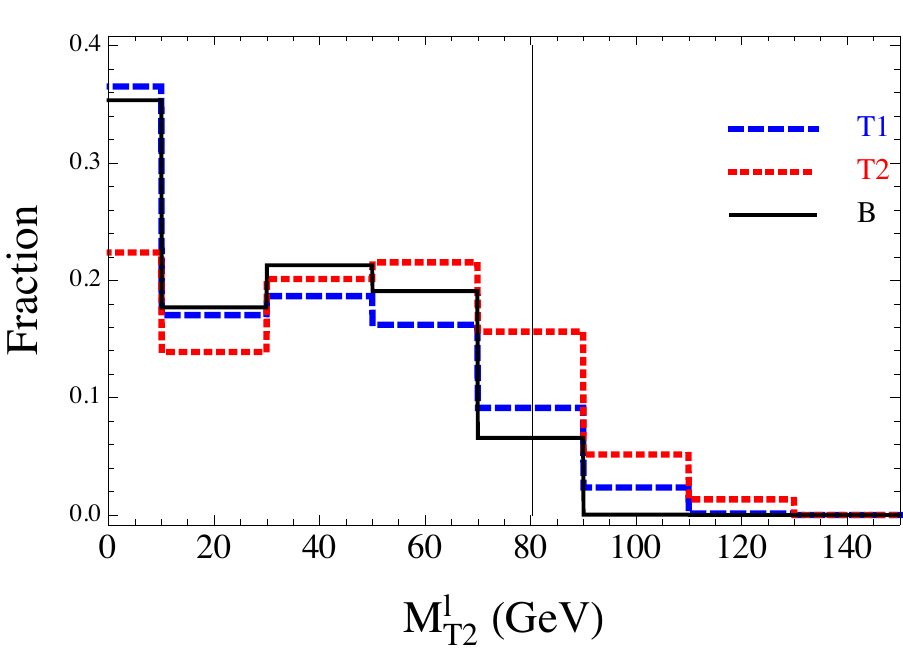}
\vspace{0.4cm} \\
\includegraphics[width=7cm]{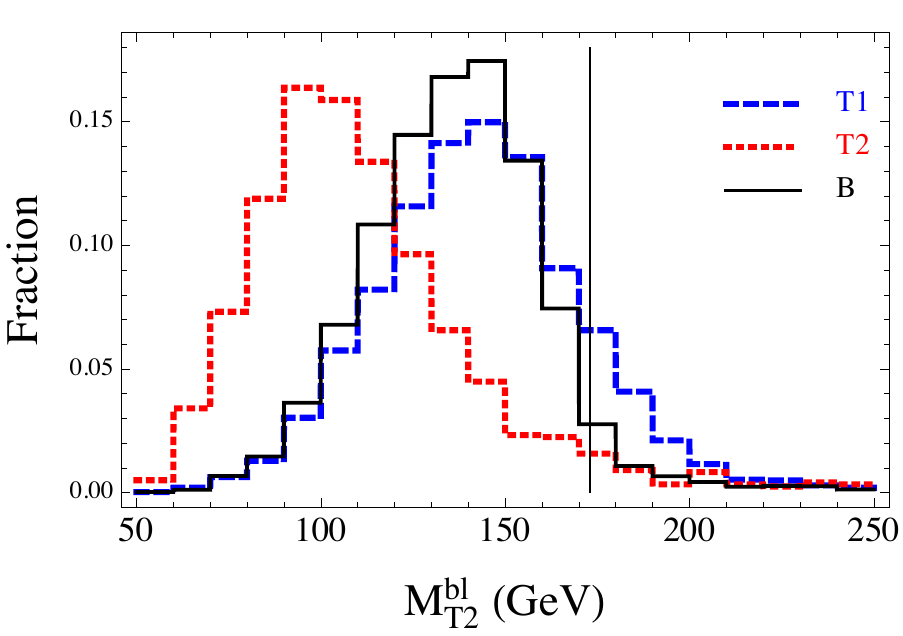}
\vspace{0.4cm} \\
\includegraphics[width=7cm]{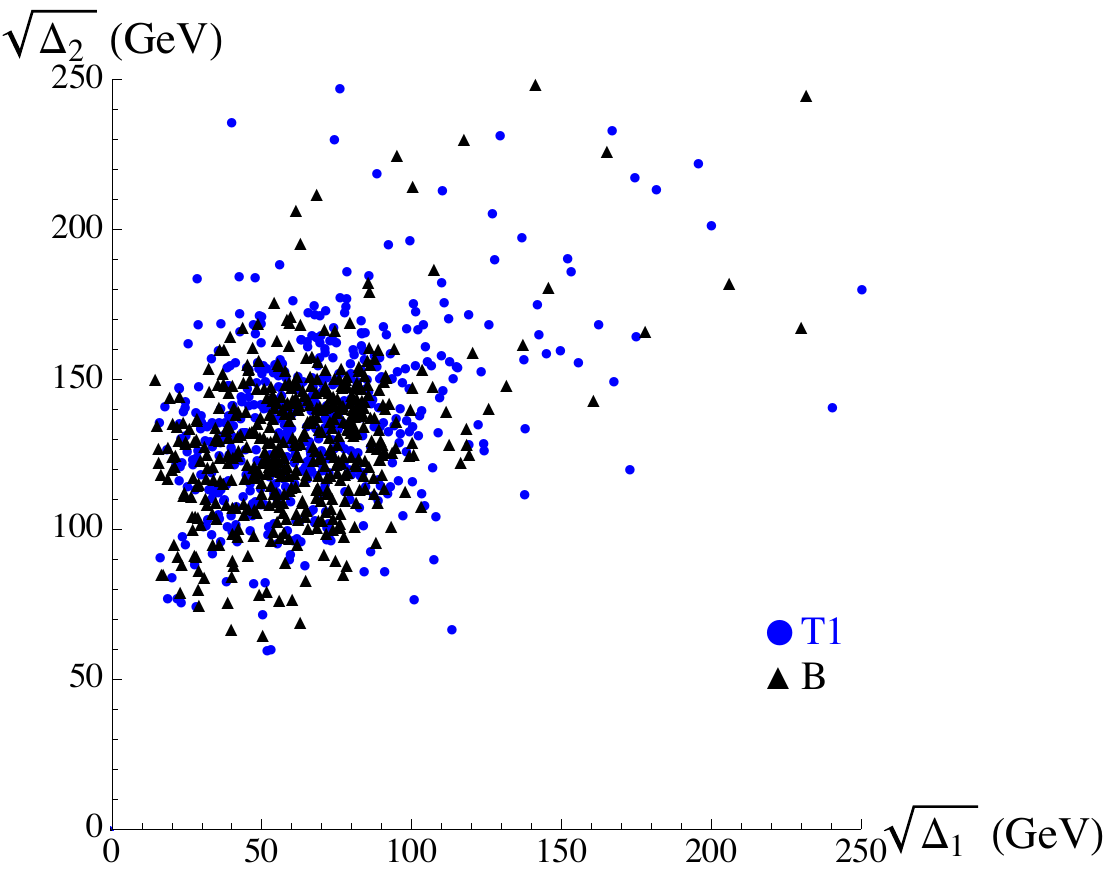} \hspace{0.4cm}
\includegraphics[width=7cm]{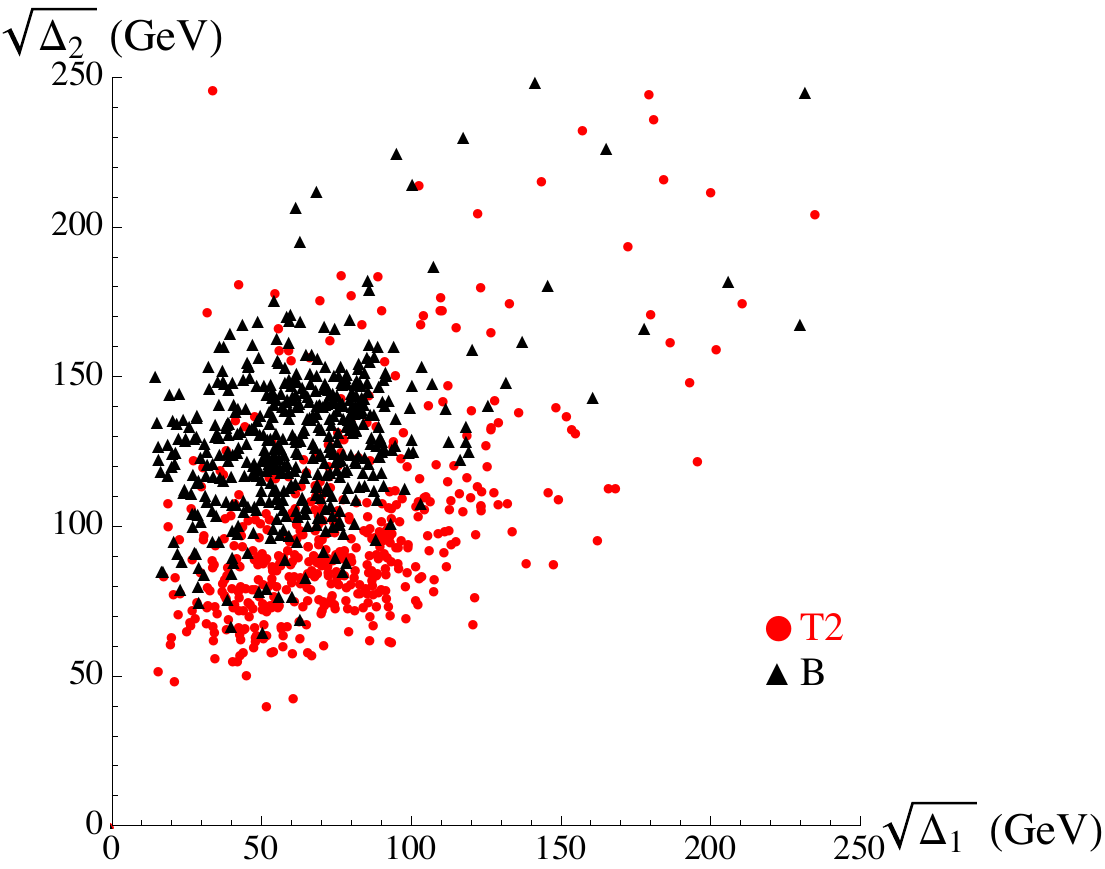}
\caption{Distributions of different variables for signal T1, T2 and $t\bar{t}$ background.  Both T1 and T2 are stop pair productions with $100\%$ branching ratio to top quark and neutralino.  The mass of the stop is $300$~GeV for T1 and $250$~GeV for T2, and the mass of the neutralino is $120$~GeV for both T1 and T2.  The distributions of T1 are very close to the ones of $t\bar{t}$ background.  The distributions of T2 are generally concentrated on smaller values than the ones of $t\bar{t}$ background, except $M^l_{T2}$, for which T2 goes beyond the endpoint of the background around $W$ mass.  The difference is mainly due to the fact that T1 has on-shell tops while the tops in T2 can only be off-shell.}
\label{fig:puretop}
\end{figure}

Before studying the case for which the stop can decay through both channels, it is useful to first look at the case of the stop decaying purely through the top quark.  We consider two spectra named T1 and T2, with the mass of the stop being $300$ ($250$) GeV for T1 (T2) and the mass of the neutralino being $120$ GeV. They correspond to S1-3 and S4-6 listed in Table~\ref{tab:spectrum} respectively.  For T1, the phase space is just enough for the stop to decay to an on-shell top and a neutralino ($\tilde{t}_1\to t \tilde{\chi}^0_1$).  For T2, the mass gap between stop and neutralino is smaller than the top mass, and the stop needs to undergo a 3-body decay via an off-shell top ($\tilde{t}_1\to b W \tilde{\chi}^0_1$).  In  Fig.~\ref{fig:puretop} we compare the distributions for several variables between T1, T2 and the $t\bar{t}$ background.  The distributions of T1 are very close to the ones of the $t\bar{t}$ background, mainly due to the presence of the on-shell tops.  For T2, the tops are off-shell, and the distributions generally shift to smaller values than the ones of $t\bar{t}$ background, except $M^l_{T2}$, for which T2 goes beyond the endpoint of the background around $W$ mass.  From these distributions, one would expect that by including this decay channel, the significance will be reduced for S1-3, while for S4-6 it is unclear.
%whether the effects are negative.  

%

\begin{figure}[ht!]
\centering
\includegraphics[width=6.7cm]{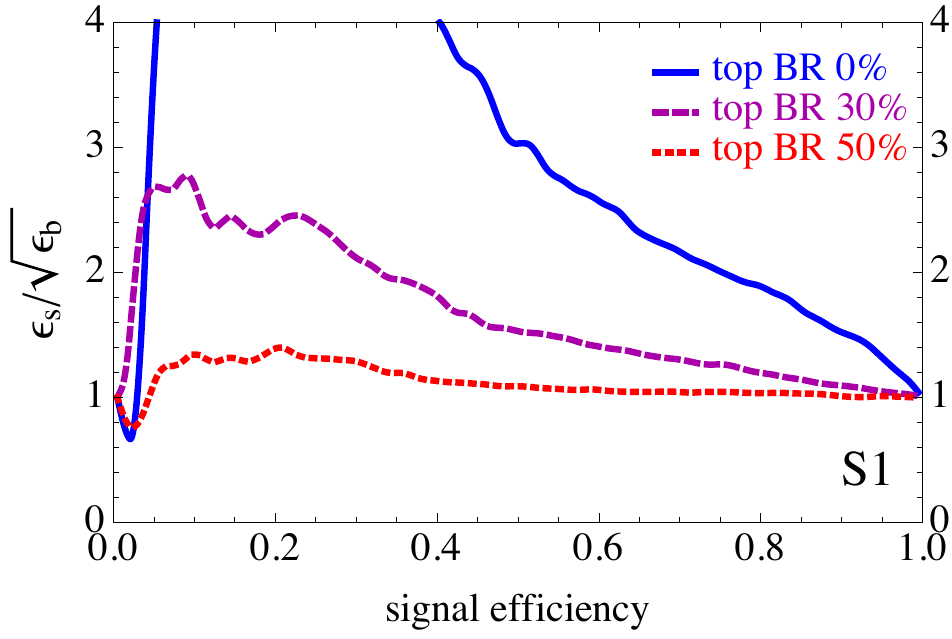} \hspace{0.7cm}
\includegraphics[width=7cm]{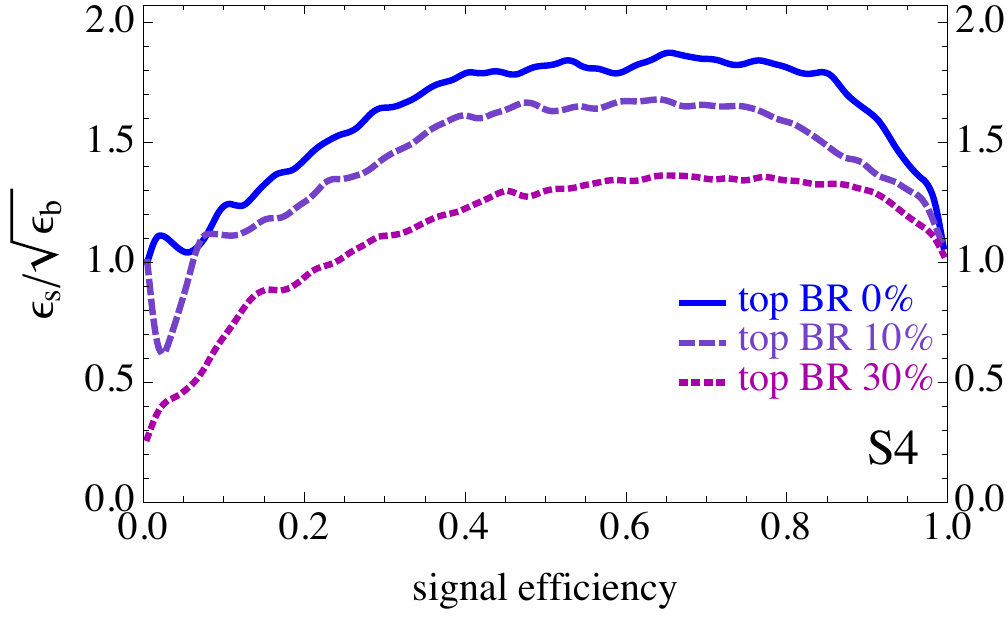}
\vspace{0.4cm} \\
\includegraphics[width=7cm]{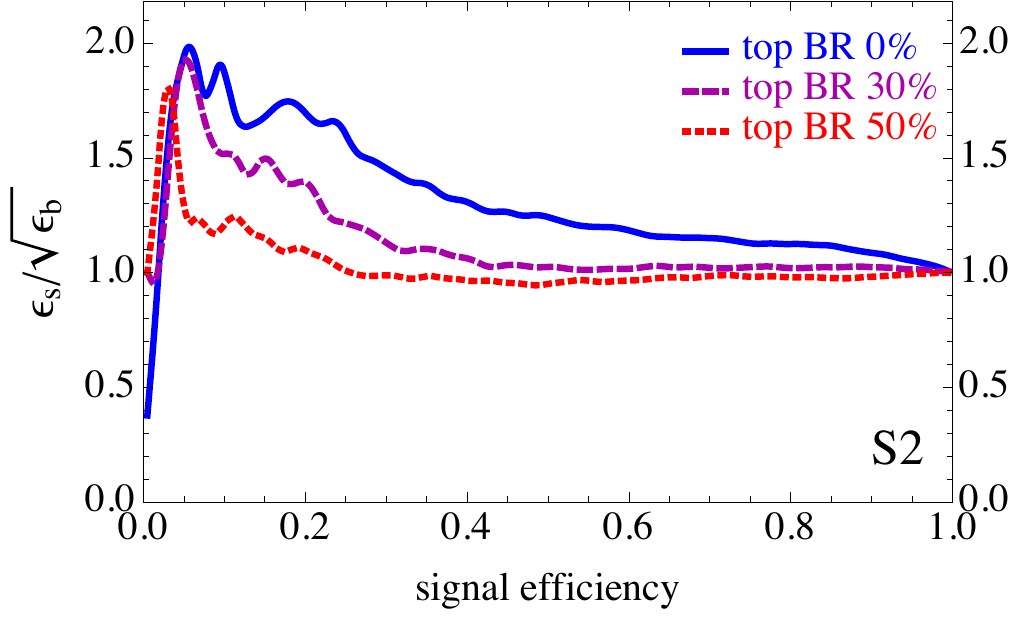} \hspace{0.4cm}
\includegraphics[width=7cm]{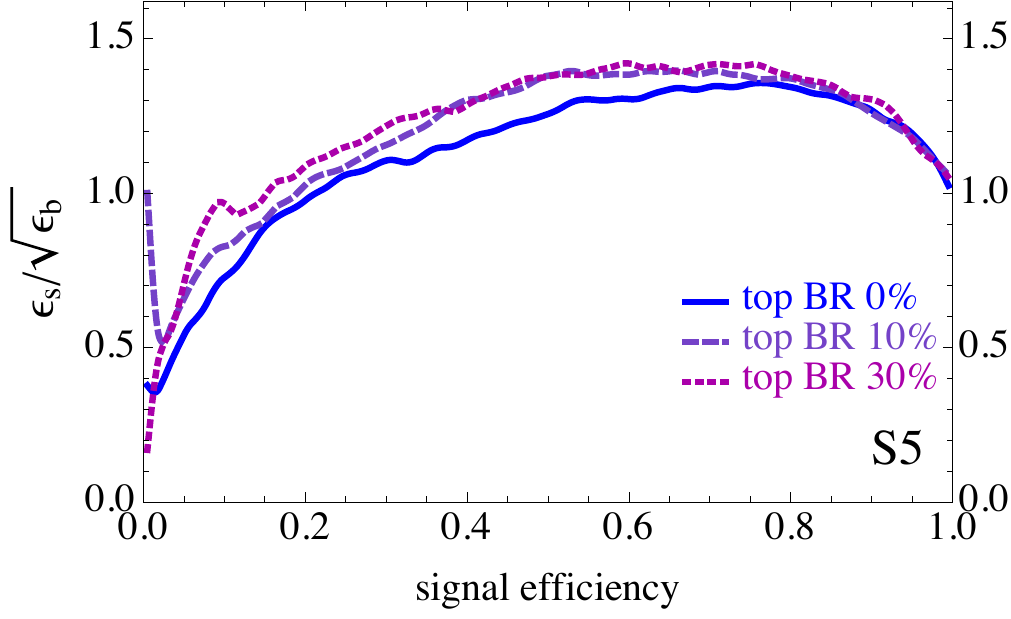}
\vspace{0.4cm} \\
\includegraphics[width=6.7cm]{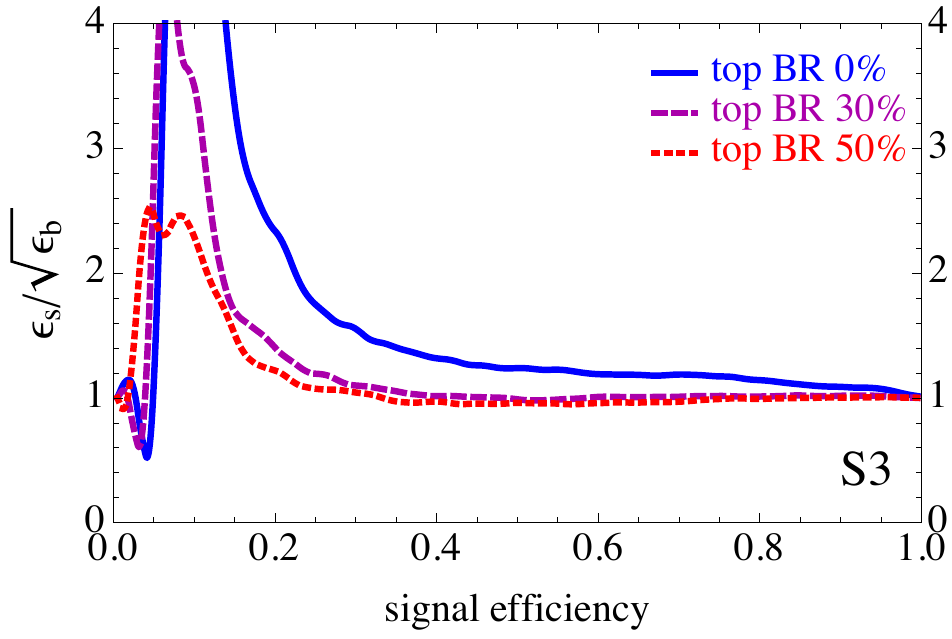} \hspace{0.7cm}
\includegraphics[width=7cm]{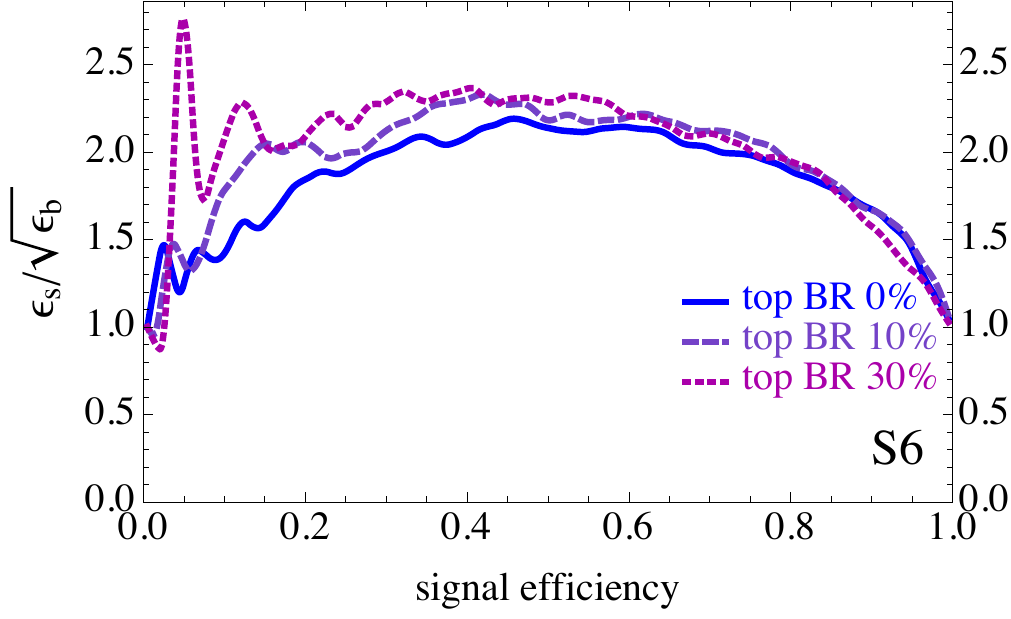}
\caption{The relative improvement on signal significance ($\epsilon_s/\sqrt{\epsilon_b}$) as a function of the signal efficiency $\epsilon_s$ after the Boost Decision Tree optimization for S1-6.  The optimizations are performed with the combination of all variables.  Three curves are shown for each signal spectrum, each assuming a different value of the branching ratio stop decaying to top and neutralino (top BR).}
\label{fig:BDTmix}
\end{figure}

For S1-3, the branching fraction of the stop decaying to a top quark and a neutralino is suppressed by the small phase space; for S4-6, it is suppressed even further by the off-shell decay.  Nevertheless, the branching ratios depend on the couplings as well as the spectrum and can be very different for different models.  In order to perform a general study, we choose a few benchmark points for the branching ratio of the stop decaying to the top quark ($30\%$ and $50\%$ for S1-3, $10\%$ and $30\%$ for S4-6) and study the effects.  The larger branching fraction benchmark points may be overestimated, and could be considered as a study of the ``worst case scenario."  The signal events contain stop pairs that 1) both decay to a chargino and a bottom quark, 2) both decay to a top quark and a neutralino, 3) one decays to a chargino and a bottom quark, the other decays to a top quark and a neutralino.  The effects can be seen in Fig.~\ref{fig:BDTmix}, which shows the relative improvement on signal significance ($\epsilon_s/\sqrt{\epsilon_b}$) as a function of the signal efficiency $\epsilon_s$ after the BDT optimization for all signal spectra S1-6, each with three different values (including $0\%$) of branching ratios to the top.  For S1-3, increasing the branching ratio to the top results a decrease in $\epsilon_s/\sqrt{\epsilon_b}$, due to the fact that signal with on-shell tops tend to look similar to the $t\bar{t}$ background.  For S4, $\epsilon_s/\sqrt{\epsilon_b}$ also decreases as the branching ratio to the top increases. The main reason is that for S4, chargino events has smaller $M^\ell_{T2}$ than the backgrounds, while top events has slightly larger $M^\ell_{T2}$ , and the resultant distribution is closer to the $t\bar{t}$ background.  For S5 and S6, $\epsilon_s/\sqrt{\epsilon_b}$ actually increases slightly as the branching ratio to the top increases.  However, $s/\sqrt{b}$ still turns out to be reduced for these two spectra, because the acceptances (after basic cuts) are lower for the three-body decay via the off-shell top.  In summary, we list $\epsilon_s/\sqrt{\epsilon_b}$, $s/\sqrt{b}$ and $s/b$ after the optimized cuts with BDT at 22~fb$^{-1}$ in Table~\ref{tab:topmix1}~and~\ref{tab:topmix2}.  To ensure the statistical reliability of the results, we require $\epsilon_s > 0.1$ and $\epsilon_b > 0.01 (0.005)$ in Table~\ref{tab:topmix1}(\ref{tab:topmix2}).

\begin{table}[bht!]
   \centering
        \renewcommand{\arraystretch}{1.3}
   \begin{tabular}{|c||c|c|c||c|c|c||c|c|c|}
     \hline \hline
& \multicolumn{3}{ |c|| }{$\epsilon_s/\sqrt{\epsilon_b}$} & \multicolumn{3}{ |c|| }{$s/\sqrt{b}$} & \multicolumn{3}{ |c| }{$s/b$} \\\hline \hline
Top BR & $0\%$ & $30\%$ & $50\%$ & $0\%$ & $30\%$ & $50\%$ & $0\%$ & $30\%$ & $50\%$      \\ \hline
S1 &  $3.98$ & $2.41$  & $1.40$ & $1.93$ & $1.38$ & $0.937$ & $0.169$ & $0.121$ & $0.057$   \\ \hline
S2 &  $1.74$ & $1.49$ & $1.21$ & $1.73$ & $1.47$ & $1.18$ & $0.154$ & $0.126$ & $0.101$     \\ \hline
S3 & $2.00$ &  $1.60$ & $1.44$ & $1.76$ & $1.41$ & $1.29$ & $0.139$ & $0.122$ & $0.106$     \\ \hline \hline
Top BR & $0\%$ & $10\%$ & $30\%$ & $0\%$ & $10\%$ & $30\%$ & $0\%$ & $10\%$ & $30\%$      \\ \hline
S4 & $1.87$ & $1.68$ & $1.36$ &  $1.44$ & $1.30$ & $1.00$ & $0.0366$ & $0.0301$ & $0.0186$    \\ \hline
S5 & $1.36$ & $1.40$ & $1.42$ & $1.65$ & $1.56$ & $1.38$ & $0.0261$ & $0.0292$ & $0.0294$     \\ \hline
S6 & $2.19$ & $2.33$ & $2.37$ & $2.60$ & $2.55$ & $2.31$ & $0.111$ & $0.127$ & $0.120$     \\ \hline \hline
   \end{tabular}
   \caption{$\epsilon_s/\sqrt{\epsilon_b}$, $s/\sqrt{b}$ and $s/b$ after the optimized cuts with BDT at 22~fb$^{-1}$, with the requirement that the signal and the background efficiencies to be $\epsilon_s \geq 0.1$ and $\epsilon_b \geq 0.01$.  The results are shown for different branching ratios of stop decaying to top and neutralino (labelled as TOP BR).} 
  \label{tab:topmix1}
\end{table}
\begin{table}[bht!]
   \centering
        \renewcommand{\arraystretch}{1.3}
   \begin{tabular}{|c||c|c|c||c|c|c||c|c|c|}
     \hline \hline
& \multicolumn{3}{ |c|| }{$\epsilon_s/\sqrt{\epsilon_b}$} & \multicolumn{3}{ |c|| }{$s/\sqrt{b}$} & \multicolumn{3}{ |c| }{$s/b$} \\\hline \hline
Top BR & $0\%$ & $30\%$ & $50\%$ & $0\%$ & $30\%$ & $50\%$ & $0\%$ & $30\%$ & $50\%$      \\ \hline
S1 &  $4.78$ & $2.46$  & $1.40$ & $2.31$ & $1.41$ & $0.937$ & $0.277$ & $0.137$ & $0.057$   \\ \hline
S2 &  $1.74$ & $1.50$ & $1.24$ & $1.73$ & $1.47$ & $1.21$ & $0.154$ & $0.171$ & $0.116$     \\ \hline
S3 & $2.50$ &  $1.76$ & $1.74$ & $2.19$ & $1.55$ & $1.56$ & $0.263$ & $0.168$ & $0.179$     \\ \hline \hline
Top BR & $0\%$ & $10\%$ & $30\%$ & $0\%$ & $10\%$ & $30\%$ & $0\%$ & $10\%$ & $30\%$      \\ \hline
S4 & $1.87$ & $1.68$ & $1.36$ &  $1.44$ & $1.30$ & $1.00$ & $0.0366$ & $0.0301$ & $0.0186$    \\ \hline
S5 & $1.36$ & $1.40$ & $1.42$ & $1.65$ & $1.56$ & $1.38$ & $0.0261$ & $0.0292$ & $0.0294$     \\ \hline
S6 & $2.19$ & $2.33$ & $2.37$ & $2.60$ & $2.55$ & $2.31$ & $0.111$ & $0.127$ & $0.120$     \\ \hline \hline
   \end{tabular}
   \caption{The same as Table~\ref{tab:topmix1}, but requiring $\epsilon_s \geq 0.1$ and $\epsilon_b \geq 0.005$.} 
  \label{tab:topmix2}
\end{table}
%

%%%%%%%%%%%%%%%%%%%%%%%%%%%%%%%%%%%%%%%%%%%%%%%%%%%%%%%

\section{Minimal Compatible Masses for Two-step Symmetric Decay Chain}
\label{app:dd}
%%%%%%%%%%%

\begin{figure}
\centering $
\begin{array}{cc}
\begin{picture}(250,100)
  \thicklines
  \put(50,100){\line(1,2){30}}
  \put(50,100){\line(1,-2){30}}
  \put(62,75){\line(1,0){40}}
  \put(102,75){\line(1,-2){17}}
  \multiput(107,75)(10,0){4}{\line(1,0){5}}
   \put(62,125){\line(1,0){40}}
  \put(102,125){\line(1,2){17}}
  \multiput(107,125)(10,0){4}{\line(1,0){5}}
   \put(45,112){$Y$}
  \put(45,83){$Y$}
  \put(63,152){$5$}
  \put(63,43){$6$}
   \put(105,152){$3$}
  \put(105,43){$4$}
   \put(80,114){$X$}
  \put(80,79){$X$}
  \put(120,114){$N$}
  \put(120,79){$N$}
  \put(132,129){$1$}
  \put(132,65){$2$}
\end{picture} &
\includegraphics[width=6cm]{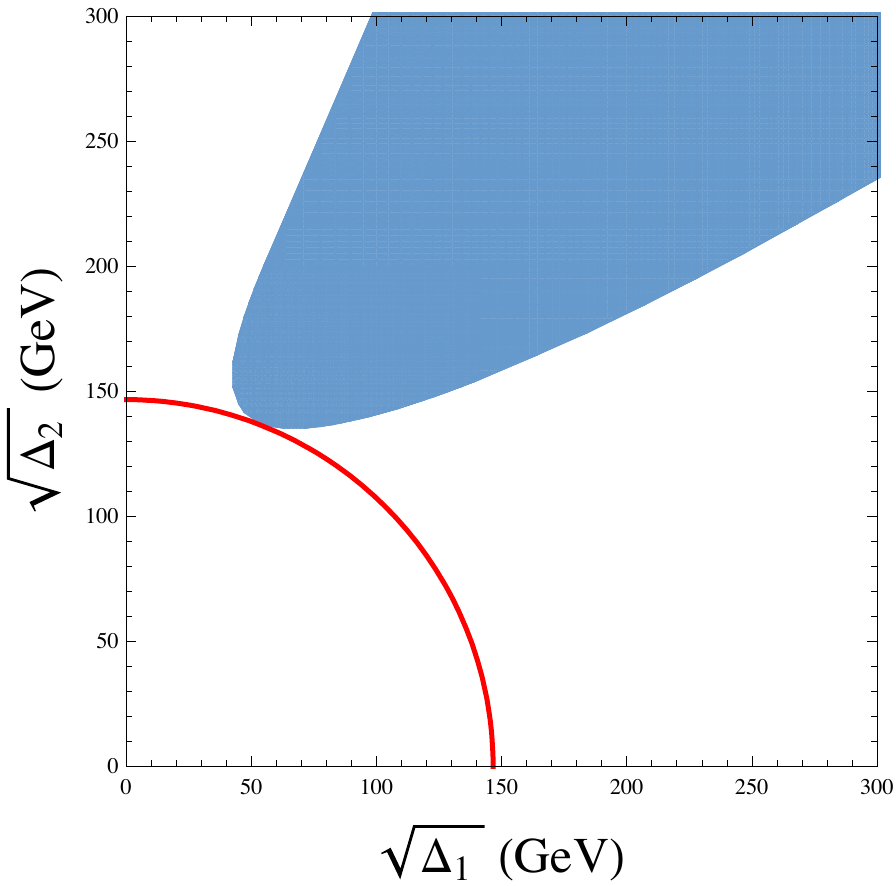}
\end{array} $
\caption{Left: Two-step symmetric decay chains.  The diagram also includes the notations that we use.  Right:  A typical allowed region (blue) in the $(\sqrt{\Delta_1}, \sqrt{\Delta_2})$ plane, where $m_N$ is set to zero, $\Delta_1 \equiv m^2_X - m^2_N$,   $\Delta_2 \equiv m^2_Y - m^2_X$.  The red curve corresponds to a constant value of $m^2_Y =  \Delta_1 + \Delta_2$.  We scan over the allowed region to find the point that gives the minimum $m_Y$.  For a fixed $m_N$, a constant $m_Y$ corresponds to a  constant $\Delta_1+\Delta_2$, which is a quarter-circle in the $(\sqrt{\Delta_1}, \sqrt{\Delta_2})$ plane.  When the quarter-circle is tangent to the compatible region (as shown by the red curve), the point of intersection gives the minimal compatible $m_Y$.}
\label{app:fig1}
\end{figure}

In this Appendix we give a detailed definition of the ``compatible-masses'' variables $\Delta_1$ and $\Delta_2$ and describe how to calculate them. We consider pair production of some particle $Y$, and each of which goes through two-step decays. It first decays to an intermediate particle $X$ plus a visible particle and then $X$ decays to an invisible particle $N$ and another visible particle. The process is shown in Fig.~\ref{app:fig1}, which also exhibits how we label the particles.  The masses of $Y$, $X$, $N$ are denoted as $m_Y$, $m_X$, and $m_N$, that will be treated as unknowns.  Particles $3$, $4$, $5$, $6$ are visible and whose 4-momenta can be experimentally measured.  For a set of test values of  $(m_Y, m_X, m_N)$, we can solve for the 4-momenta of missing particles 1 and 2 when $Y$, $X$, $N$  are on-shell.  To see this, let us first focus on one decay chain (e.g., the chain with particles $1$, $3$, and $5$).  The mass shell conditions are
\begin{equation}
p^2_1=m^2_N, \hspace{1cm} (p_1+p_3)^2=m^2_X, \hspace{1cm} (p_1+p_3+p_5)^2=m^2_Y.
\end{equation}
Taking the differences between the mass shell equations we have
\begin{eqnarray}
2p_1p_3 &=& m^2_X - m^2_N -p^2_3 \equiv \Delta_1 -p^2_3 \equiv \Delta'_1, \\
2p_1p_5 &=& m^2_Y - m^2_X -p^2_5 - 2p_3p_5 \equiv \Delta_2-p^2_5 - 2p_3p_5 \equiv \Delta'_2,
\end{eqnarray}
where 
\begin{eqnarray}
\Delta_1  &\equiv& m^2_X - m^2_N, \\
\Delta_2  &\equiv& m^2_Y - m^2_X,
\end{eqnarray}
and we further define $ \Delta'_1$ and $\Delta'_2$ to simplify the expressions.  

Expanding the 4-momenta the equations can be written explicitly as
\begin{eqnarray}
E_1E_3-p_{1x}p_{3x}-p_{1y}p_{3y}-p_{1z}p_{3z} &=& \frac{\Delta'_1}{2},  \label{b1}\\
E_1E_5-p_{1x}p_{5x}-p_{1y}p_{5y}-p_{1z}p_{5z} &=& \frac{\Delta'_2}{2}. \label{b2}
\end{eqnarray}
Combining these two equation we can eliminate $E_1$ and express $p_{1z}$ in terms of $p_{1x}$ and $p_{1y}$, 
\begin{equation}
p_{1z}=A\,p_{1x}+B\,p_{1y}+C,   \label{b3}
\end{equation}
where
\begin{equation}
A \equiv \frac{E_3p_{5x}-E_5p_{3x}}{E_5p_{3z}-E_3p_{5z}}, \hspace{1cm} B \equiv \frac{E_3p_{5y}-E_5p_{3y}}{E_5p_{3z}-E_3p_{5z}}, \hspace{1cm} C \equiv \frac{E_3\Delta'_2-E_5\Delta'_1}{2(E_5p_{3z}-E_3p_{5z})}. \label{b4}
\end{equation}
Substituting $E_1=\sqrt{m^2_N+p^2_{1x}+p^2_{1y}+p^2_{1z}}$ and Eq.~(\ref{b3}) back into Eq.~(\ref{b2}), we obtain a quadratic equaion:
\begin{equation}
a~p^2_{1x} + 2b~p_{1x}p_{1y} + c~p^2_{1y} + 2d~p_{1x} +2f~p_{1y}+g  = 0, \label{b5}
\end{equation}
where
\begin{eqnarray}
a&=&E^2_5(1+A^2)-(p_{5x}+p_{5z}A)^2 ,   \nonumber\\
b&=&E^2_5AB-(p_{5x}+p_{5z}A)(p_{5y}+p_{5z}B) ,  \nonumber\\
c&=&E^2_5(1+B^2)-(p_{5y}+p_{5z}B)^2 ,  \nonumber\\
d&=&E^2_5 AC - (p_{5x}+p_{5z}A)(p_{5z}C+\frac{\Delta'_2}{2}) ,  \nonumber\\
f&=&E^2_5 BC - (p_{5y}+p_{5z}B)(p_{5z}C+\frac{\Delta'_2}{2}) ,  \nonumber\\
g&=&E^2_5(m^2_N+C^2)-(p_{5z}C+\frac{\Delta'_2}{2})^2,
\end{eqnarray}
and $A$, $B$ and $C$ are defined in Eq.~(\ref{b4}).  The quadratic equation (\ref{b5}) describes an ellipse in the $(p_{1x},p_{1y})$ plane.

We can do the same for the other chain and obtain a similar equation in terms of $p_{2x}$ and $p_{2y}$, which also describes an ellipse in the $(p_{2x},p_{2y})$ plane.  We can then use the knowledge of the total missing transverse momentum ($\vec{\not \! p}_T = \vec{p}_{1T} + \vec{p}_{2T}$) to put both ellipses on the same $(p_{1x},p_{1y})$ plane, which are describe by two quadratic equations: 
\begin{eqnarray}
a_1~x^2 + 2b_1~xy + c_1~y^2 + 2d_1~x+2f_1~y+g_1  &=& 0\,,  \\
a_2~x^2 + 2b_2~xy + c_2~y^2 + 2d_2~x +2f_2~y+g_2  &=& 0\,.
\end{eqnarray}
where for notation simplicity we have written $(p_{1x},p_{1y})$ as $(x,y)$.  The coefficients are functions of $m_Y$, $m_X$ and $m_N$ (or equivalently, $\Delta_1$, $\Delta_2$ and $m_N$).  The two quadratic equations render 4 solutions in general which may be complex.  In order to have physical (real) solutions for the momenta, the two ellipses must intersect.

For a given event, one can check whether a set of trial masses $(\Delta_1,\Delta_2, m_N)$ is compatible by testing whether the two ellipses intersect.  This can be done efficiently without solving the actual equations by using the Sturm sequence~\cite{sturm}, similar to the calculation of $M_{T2}$ in Ref.~\cite{Cheng:2008hk}.  If we fix $m_N=0$ (which is a good assumption for the $t\bar{t}$ di-leptonic background since the missing particles are neutrinos), we have two parameters $\Delta_1$ and $\Delta_2$ left and we can scan over the $(\Delta_1, \Delta_2)$ plane to find the compatible region with that particular event, i.e., any point outside the compatible region does not have physical solution for the kinematics.  A typical compatible region for an event is shown in Fig.~\ref{app:fig1}.  There are in principle multiple ways to extract useful information or variables from the compatible region. We consider a simple projection to a pair of variables which are the coordinates of the point at the tip (with minimum $\Delta_1+\Delta_2=m_Y^2-m_N^2$) of the compatible region. It can be obtained by scanning  over the allowed region to find the point that gives the minimum  $m^2_Y$.  We output the coordinates $(\Delta_1, \Delta_2)$ of that point as our variables. Equivalently, we can also use the radial coordinates $m_Y^2$ and $\tan^{-1} (\Delta_2/\Delta_1)$ as the independent variables.

For the $t\bar{t}$ di-leptonic background, particles $Y$, $X$, $N$ correspond to the top, $W$-boson and neutrino.  The signal does not have the correct topology for this set of variables. Nevertheless, since it has the same signature as the background (2 lepton + 2 $b$-jets +$\met$), we can calculate these variables anyway and obtain a set of output values $(\Delta_1, \Delta_2)$ for each event.  Even for the wrong topology, these variables still have a strong correlation with the hardness of the $b$-jets and leptons.

A detailed description of the calculation of $(\Delta_1, \Delta_2)$ including how to test whether two ellipses intersect, and the computer codes for calculating them can be found at the following website: \\ \texttt{https://sites.google.com/a/ucdavis.edu/mass/}.

%%%%%%%%%%%%%%%%%%%%%%%%%%%%%%%%%%%%%%%%%%%%%%%%%%%
\providecommand{\href}[2]{#2}\begingroup\raggedright\endgroup

\end{document}